\documentclass[12pt]{article} 

\usepackage{url,hyperref,lineno,microtype,subcaption}
\usepackage[onehalfspacing]{setspace}

\usepackage{amsmath,amsfonts,amssymb,graphicx,amsthm,geometry}
\usepackage{stackrel,stmaryrd,mathabx,etoolbox,framed,accents}
\usepackage[all]{xy}

\def\be{\begin{equation}}
\def\ee{\end{equation}}
\def\ba{\begin{eqnarray}}
\def\ea{\end{eqnarray}}

\makeatletter
\newsavebox{\@brx}
\newcommand{\llangle}[1][]{\savebox{\@brx}{\(\m@th{#1\langle}\)}%
  \mathopen{\copy\@brx\kern-0.5\wd\@brx\usebox{\@brx}}}
\newcommand{\rrangle}[1][]{\savebox{\@brx}{\(\m@th{#1\rangle}\)}%
  \mathclose{\copy\@brx\kern-0.5\wd\@brx\usebox{\@brx}}}
\makeatother

\newgeometry{vmargin={25mm}, hmargin={22mm,22mm}} 

\begin{document}

\setcounter{tocdepth}{1}

\title{Geometric tool kit for higher spin gravity (part II):\\\vspace{2mm} An introduction to Lie algebroids\\ and\\ their enveloping algebras} 

\author{Xavier Bekaert}

\date{Institut Denis Poisson, Unit\'e Mixte de Recherche $7013$ du CNRS\\
Universit\'e de Tours, Universit\'e d'Orl\'eans\\
Parc de Grandmont, 37200 Tours, France\\
\vspace{2mm}
{\tt xavier.bekaert@lmpt.univ-tours.fr}
}

\maketitle

\vspace{5mm}

\begin{abstract}
These notes provide a self-contained introduction to Lie algebroids, Lie-Rinehart algebras and their universal envelopes. This review is motivated by the speculation that higher-spin gauge symmetries should admit a natural formulation as enveloping algebras of Lie algebroids since rigid higher-spin algebras are enveloping algebras of Lie algebras. 
Nevertheless, the material covered here may be of general interest to anyone interested in the description of gauge symmetries, connections and covariant derivatives, in terms of Lie algebroids.
In order to be self-contained, a concise introduction to the algebraic characterisation of vector bundles as projective modules over the algebra of functions on the base manifold is provided.
\end{abstract}

\thispagestyle{empty}

\pagebreak

\setcounter{page}{0}

\setcounter{tocdepth}{2}

\tableofcontents

\pagebreak

\section{Introduction}

One slogan for motivating algebroids for physicists could be: ``\textit{Lie algebroids are to local symmetries what Lie algebras are to global symmetries}''. In other words, Lie algebroids\footnote{The classical textbooks on Lie algebroids are by Mackenzie \cite{Mackenzie,Mackenzie2}. Some textbooks on related subjects include a review chapter devoted to Lie algebroids, see \textit{e.g.} \cite[Chap.6]{Moerdijk:2003} or \cite[Chap.1]{CrampinSaunders}. Idem for lecture notes, see \textit{e.g.} \cite[Lect.2]{Crainic:2011}.} are the natural tool for the mathematical description of gauge symmetries. But there is more: Lie algebroids also provide the modern framework for the theory of connections (and their generalisations).\footnote{\label{McK}The recognition that ``infinitesimal connection theory -- the part of standard connection theory which does not depend on the concepts of path-lifting or holonomy -- can be developed entirely within the concept of transitive Lie algebroids'' \cite[p.185]{Mackenzie} is one of the seminal insights of Mackenzie, which definitely established the importance of transitive Lie algebroids in differential geometry.}
In a sense, which can be made precise (and which will be reviewed here), the theory of Lie algebroid representations (respectively, extensions) coincides with the theory of flat (respectively, curved) connections on principal and vector bundles.\footnote{Strictly speaking, this is true for ``infinitesimal connection theory'' in the sense of footnote \ref{McK}. In any case, the part of connection theory which \textit{does} depend on the concepts of path-lifting or holonomy can be developed within the realm of Lie groupoids (which are to Lie algebroids what Lie groups are to Lie algebras).}  

\vspace{2mm}

An important companion of any Lie algebra is its universal enveloping algebra, which can be thought as its natural counterpart as an associative algebra. It is remarkable that this algebraic construction admits a generalisation for Lie algebroids. Geometrically speaking, this construction amounts to switch from vector fields to differential operators. In physicist jargon, such a step amounts to switch from usual spacetimes symmetries to higher-spin symmetries. Accordingly, higher-spin gravity provides a motivation for studying the enveloping algebras of Lie algebroids.

\vspace{2mm}

Higher-spin gravity\footnote{Many pedagogical reviews of various levels are available by now: advanced ones \cite{Reviews} as well as introductory ones \cite{Introductions}. Two books of conference proceedings also offer a panorama of this research area \cite{Proceedings}.} is formulated, in the frame-like approach, in terms of a Cartan-like connection: a differential one-form on the spacetime manifold taking values in the higher-spin algebra. The modern definition of higher-spin algebras is as enveloping algebras of rigid spacetime symmetries (isometries or conformal transformations).\footnote{More precisely, a higher-spin algebra is the quotient of the universal enveloping algebra of the corresponding spacetime symmetry algebra by the annihilator of a suitable module over the latter algebra.} Generalising to higher spins our slogan (``Lie algebroids are to local symmetries what Lie algebras are to global symmetries'') suggests that the natural geometrical tool for properly addressing the gauging of rigid higher-spin symmetries are enveloping algebra of Lie algebroids.
In down-to-earth terms, the point is that higher-spin gauge symmetries should be suitable subalgebras of differential operators on a principal bundle (say the frame bundle underlying the standard gravity that the higher-spin gravity extends). However, this description is somewhat obscured in standard treatments (for instance, the identification of higher-spin local Lorentz transformations is complicated at nonlinear level). One hope is that universal enveloping algebras of Lie algebroids might help to address this issue.
One related hope is that enveloping algebras of Lie algebroids might also help improving our geometrical understanding of higher-spin Cartan-like connections.\footnote{Note that the description of connections as decompositions of Lie algebroids as (semidirect) sums, admits an associative analogue \cite{Bekaert:2022dlx} as factorisations of universal enveloping algebras of Lie algebroids as (smashed) products. A reformulation of higher-spin Cartan-like connections in similar terms remains a tantalising dream.} This expectation is not unreasonable since, for usual gravity, it is by-now well-established that a natural framework for formulating Cartan connections is precisely the language of Lie algebroids (see \textit{e.g.} \cite{CrampinSaunders,Crampin}).

\vspace{2mm}

The classical textbooks \cite{Mackenzie,Mackenzie2} by Mackenzie on Lie algebroids provide a thorough review of the subject, but universal enveloping algebras are not covered there. To the best knowledge of the author, this material is absent from any existing review papers, although brief pedagogical reviews can be found in various research papers (see \textit{e.g.} \cite[Section 1.4]{Bekaert:2022dlx} or \cite[Section 2]{Moerdijk}). The goal of the present notes is to partially fill this gap by providing a self-contained introduction to the universal enveloping algebras of Lie-Rinehart algebras for non-specialists.

\vspace{4mm}

The following subsection draws up the plan of these notes.
It is followed by a subsection providing a brief and non-technical explanation of the sense in which the theory of Lie algebroid representations/extensions coincides with the theory of flat/curved connections on principal and vector bundles. The introduction ends with a short subsection containing minor preliminary remarks. 

\subsection{Plan}

Section \ref{vectbdlsprojmods} provides a short introduction to the profound  reformulation of vector bundles as projective modules over the structure algebra (i.e. the commutative algebra of functions on the base manifold), known as the Serre-Swan theorem. 

This material is useful in order to formulate, in Section \ref{LARalgebras}, Lie algebroids as Lie-Rinehart algebras over the structure algebra of a smooth manifold.
Many examples of Lie-Rinehart algebras and Lie algebroids are provided, as well as an extensive review of the formulation of linear and principal connections in this language. 

Finally, in Section \ref{Assalgebras} various equivalent definitions and descriptions of the universal enveloping algebra of a Lie-Rinehart algebra (in particular of a Lie algebroid) are reviewed in detail.

\subsection{Infinitesimal connections as decompositions of Lie algebroids: a brief sketch}

It was advertised above that, in a sense the theory of Lie algebroid representations (respectively, extensions) coincides with the theory of flat (respectively, curved) connections on principal and vector bundles.
These ideas can be summarised (or, rather, sketched) as follows.

Remember that the automorphisms of a fibred manifold are those diffeomorphisms that preserve the fibration (and the potential extra structure on the bundle, \textit{e.g.} the group action for a principal bundle or the linear structure for a vector bundle), \textit{i.e.} they map fibres to fibres. These automorphisms form an infinite-dimensional group. The latter groups are notoriously cumbersome: hardcore analysis is required to endow them with a Lie group structure (some topology should carefully be chosen, etc). One can circumvent this obstacle by focusing on their infinitesimal counterparts. Infinitesimal automorphisms can be seen as vector fields on the bundle preserving the underlying structure. As is well-known, vector fields can be thought as infinitesimal diffeomorphisms since they generate one-parameter groups of diffeomorphisms (aka flows on the manifold). In particular, those vector fields that generate flows of automorphisms can be interpreted as infinitesimal automorphisms of the fibre bundle (\textit{e.g.} invariant vector fields for principal bundles or covariant derivatives for vector bundles). They span an algebra which has a rich structure: it is a Lie-Rinehart algebra. In the case of principal and vector bundles, this structure is even richer and defines a Lie algebroid: the Atiyah algebroid of the corresponding bundle.
 
From the above point of view, a connection consists in a device lifting infinitesimal diffeomorphisms of the base manifold (aka vector fields thereon) to infinitesimal automorphisms of the total space of the fibre bundle, in a way compatible with the fibration. This compatiblity condition only takes into account some part of the algebraic structure\footnote{Technically speaking, only the bimodule structure of the algebra of first-order differential operators on the base is taken into account.} of vector fields: it does not take into account its Lie algebra structure when the connection is curved. Indeed, if the lift of infinitesimal symmetries preserves the full Lie-Rinehart algebra structure, then the connection is flat. In this sense, flat connections identify with faithful representations of the algebra of infinitesimal diffeomorphisms of the base manifold where the latter are represented as some infinitesimal automorphisms of the fibre bundle.
Consequently, a flat connection is equivalent to a decomposition of the algebra of infinitesimal automorphisms of the fibre bundle as a semidirect sum of the algebra of infinitesimal diffeomorphisms of the base manifold and the algebra of vertical automorphisms  of the fibre bundle (\textit{e.g.} invariant vertical vector fields for principal bundles or infinitesimal changes of frames for vector bundles) where the former algebra acts on the latter. This algebraic definition allows for a natural generalisation where a curved connection is defined as a  similar decomposition as a semidirect sum. In Lie algebra theory, such decompositions are called split extensions.
It is in this sense that the theory of Lie algebroid representations (respectively, extensions and their splittings) coincides with the theory of flat (respectively, curved) connections.

\subsection{Minor remarks}

This paper is the second part of a series to appear (see the first part \cite{Bekaert:2023cmi} for the disclaimer and teaser, which remain in order). Notations and terminology have been selected for consistency with the other parts in the series. I present my apologies to my fellow physicists because there will be very few formulas written in coordinates and in components (but this will be compensated in the third part of the series where there will be plenty of these).
Substantial amount of notions from abstract algebra are necessary to describe properly the structure of infinitesimal symmetries of fibre bundles in terms of Lie-Rinehart algebras. 

\vspace{1mm}

One of the advertisements at the beginning was the observation (nowadays almost a platitude) that Lie algebroids are a natural tool for the mathematical description of gauge symmetries. This truism should be clarified in order to avoid confusion since there are (at least) two possible interpretations of this informal statement. They should be carefully distinguished. On the one hand, a first interpretation of this truism is that infinitesimal gauge symmetries are sections of a vector bundle (with finite rank if the gauged Lie group is finite-dimensional) endowed with a structure of transitive Lie algebroid. More precisely, this is the Atiyah bundle instrumental for formulating (Cartan, linear, principal, etc) connections in the language of Lie algebroids. 
On the other hand, a second interpretation of the truism is that gauge theories can always be formulated in terms of Lie-$\infty$ structures and Q-manifolds. Lie algebroids are a particular example of the latter. However, in this case the Lie algebroid structure is in target space (``field space'' or, to be more precise, ``jet space'') which is typically infinite-dimensional for field theories with local degrees of freedom. This second incarnation of the deep relevance of Lie algebroids for gauge theories is an extremely vast and beautiful area but, to be clear, it will not be touched at all here.

\pagebreak

\section{Vector bundles as modules over the structure algebra}
\label{vectbdlsprojmods}

The algebraic reconstruction of a manifold from its structure algebra (see \textit{e.g.} \cite[Sect.2]{Bekaert:2023cmi} or \cite[Chap.3-4]{Nestruev}) has a suitable generalisation for fibre bundles (see \textit{e.g.} \cite[Chap.11-12]{Nestruev}), which will be very useful in the next sections. In order to phrase this algebraic reconstruction, several basic notions of module theory\footnote{The reader is referred to textbooks on modern algebra (see \textit{e.g.} \cite[Section 7.4]{Rotman}) for more details and proofs.} are needed.

\subsection{Lost in translation}

Some frightening words of modern mathematics literature are briefly introduced in the present subsection because they will be used repeatedly throughout the notes. They provide such a convenient tool, for organising material and summarising many concepts in few words, that we decided to take the risk of using them, despite the fact it might scare some colleagues.

\paragraph{Short exact sequences.}
A short exact sequence of morphisms of vector spaces is a succession of four linear maps, 
\be
0\to A\stackrel{i}{\hookrightarrow}B\stackrel{\pi}{\twoheadrightarrow}C\to 0\,,\label{shortexact}
\ee
between the vector spaces $A$, $B$, $C$ and the trivial vector space denoted $0$, such that the image of each map coincides with the kernel of the following map.
For all practical purpose, the short exact sequence \eqref{shortexact} is a pictorial way to summarise the isomorphism $C\cong B/A$ between the image $C$ (of the surjective morphism $\pi$) and the quotient $B/A$ (of the source $B$ of the surjective morphism $\pi$, by the kernel $A$ of the injective morphism $i$).
A splitting of the short exact sequence \eqref{shortexact} is a reversed short exact sequence
\be
0\leftarrow A\stackrel{r}{\twoheadleftarrow}B\stackrel{s}{\hookleftarrow}C\leftarrow 0\,,\label{splitting}
\ee
where $r\circ i=id_A$ and $\pi\circ s=id_C$ (\textit{i.e.} the map $r$ is a retraction of the injection $i$ while the map $s$ is a section of the surjection $\pi$).
The splitting \eqref{splitting} of the short exact sequence \eqref{shortexact} expresses that $B= i(A)\oplus s(C)$. In fact, one has the relation
$i\circ r+s\circ\pi=id_B\,$. Short exact sequences and splittings rarely belong to the mathematical toolkit of the common theoretical physicist. However, as emphasised by Mackenzie, they provide a unifying and extremely concise description of connections. This modern viewpoint on the theory of connections is by now common lore among differential geometers. The use of these basic tools in a variety of context will be one of the leitmotiv of these notes, hoping to contribute popularising this useful conceptual language among colleagues. (For the sake of simplicity, one will often abuse terminology and notation later on by treating $A$ directly as a subspace of $B$ and note accordingly $A\subset B$ although, strictly speaking, one should write $i(A)\subset B$.)

\paragraph{Modules.} Let $\mathcal A$ be an algebra (either Lie or associative here). The term $\mathcal A$-module $V$ will be used as a shorter synonym for ``representation space $V$ of the algebra $\mathcal A$''.\footnote{Note that when $\mathcal A$ is a field, say $\mathbb K$, then a $\mathbb K$-module $V$ provides a shorter synonym for ``vector space $V$ over the field $\mathbb K$''. In this case, one will nevertheless stick to the usual terminology ``vector space'' because the field will be left implicit most of the time.} Incidentally, the term ``submodule" also provides a brief synonym for ``invariant subspace''. The algebra $\mathcal A$ can act from the left or the right, so one respectively speaks of left or right $\mathcal A$-modules. An $\mathcal A$-bimodule is both a left and right $\mathcal A$-module, and moreover the left and right action commute with each other.
If all morphisms of the short exact sequence \eqref{shortexact} stand for morphisms of $\mathcal A$-modules, then the short exact sequence means that $A$ is a submodule of $B$ and that $C$ is the quotient module. 
When $\mathcal A$ is a commutative algebra, morphisms of $\mathcal A$-modules identify with $\mathcal A$-linear maps.
The space of $\mathcal A$-linear maps from the $\mathcal A$-module $V$ to the $\mathcal A$-module $W$ is denoted $\text{Hom}_{\mathcal A}(V,W)$.
When $\mathcal A$ is a field $\mathbb K$, one drops the explicit mention of $\mathbb K$ whenever possible. For instance, the space of linear maps from a vector space $V$ to a vector space $W$ is denoted $\text{Hom}(V,W)$ and, in particular, $\text{Hom}(V,\mathbb{K})$ is denoted $V^*$.

\subsection{Free modules over commutative algebras}

Let ${\mathcal A}$ be an algebra over the field $\mathbb{K}$.
\vspace{2mm}

An ${\mathcal A}$-module $\cal M$ is generated by a subset ${\mathcal B}\subset{\mathcal M}$ if ${\mathcal M}=\sl{span}_{\mathcal A}\,{\mathcal B}$, \textit{i.e.} all elements of $\cal M$ can be written as linear combinations of elements of $\cal B$ with coefficients in $\mathcal A$.\footnote{An implicit, but crucial, assumption is that the sum of terms is finite, \textit{i.e.} only a finite number of coefficients are non-vanishing.} In this case, the subset $\cal B$ is called a \textbf{generating set} of the ${\mathcal A}$-module $\cal M$.
When there exists a generating set with a finite number of elements, the ${\mathcal A}$-module $\cal M$ is said \textbf{finitely-generated}.

A generating set of linearly independent elements is called a \textbf{basis}. 
An ${\mathcal A}$-module $\cal M$ possessing a basis $\cal B$ is said \textbf{freely-generated} and called a \textbf{free module}.
The number of elements in a basis is called the \textbf{rank}, but it may depend
on the choice of the basis (see \textit{e.g.} \cite[Example 8.22]{Rotman}).
An algebra  $\mathcal A$  for which the rank of free  $\mathcal A$-modules does not depend on the choice of the basis
is usually said to satisfy the ``invariant basis number'' property. For instance, any commutative
algebra has this property (see \textit{e.g.} \cite[Proposition 7.50]{Rotman}).

\vspace{3mm}
\noindent{\small\textbf{Example (Vector spaces as modules over a field)\,:} Every module over a field $\mathbb K$ is free. In fact, $\mathbb K$-modules identify with vector spaces over $\mathbb K$.}
\vspace{3mm}

Any module is isomorphic to the quotient of a free module (see \textit{e.g.} \cite[Proposition 7.51]{Rotman}). In other words, every module has a presentation by generators and relations. In particular, one can show that a finitely-generated module is a quotient of a free module with a finite basis.

\vspace{3mm}
\noindent{\small\textbf{Example (Free module as a tensor product)\,:} If a free ${\mathcal A}$-module $\cal M$ and a $\mathbb K$-module $V$ (\textit{i.e.} a vector space) possess a common basis (say $\cal B$, hence ${\mathcal M}=\sl{span}_{\mathcal A}\,{\mathcal B}$ and $V=\sl{span}_{\mathbb K}\,{\mathcal B}$) then ${\mathcal M}\cong{\mathcal A}\otimes V$ as a vector space over $\mathbb K$.
}
\vspace{3mm}

The \textbf{tensor algebra} $\otimes_{\mathcal A}({\mathcal M})$ \textbf{of the module} $\cal M$ \textbf{over the commutative algebra} $\mathcal A$ is the free associative algebra generated by monomials in $\cal M$ with commuting coefficients in $\mathcal A$. 

\vspace{3mm}
\noindent{\small\textbf{Example (Tensor algebra of a vector space)\,:} The tensor algebra $\otimes(V)$ of a $\mathbb K$-module $V$  is the free associative algebra spanned by linear combinations of monomials in vectors of $V$. Let ${\mathcal B}=\{e_i\,|\,i\in I\}$ be a basis of the vector space $V$. Then the infinite set $\{e_{i_1}\otimes\cdots\otimes e_{i_k}\,\,|\,\,i_1,\cdots, i_k\in I\,\,\text{and}\,\,k\in\mathbb{N}\}$ provides a basis of the tensor algebra $\otimes(V)$.
}

\vspace{3mm}
\noindent{\small\textbf{Example (Tensor algebra of a free module)\,:} The tensor algebra $\otimes_{\mathcal A}({\mathcal M})$ of a free $\mathcal A$-module ${\mathcal M}={\mathcal A}\otimes V$, generated by the vector space $V$, is isomorphic
to the free $\mathcal A$-module generated by the tensor algebra of the vector space $V$ over the underlying field:
\be
\bigotimes_{\mathcal A}\big({\mathcal A}\otimes V\big)\,\cong\,{\mathcal A}\,\otimes\,\bigotimes(V)\,.
\ee
In fact, since the free ${\mathcal A}$-module $\cal M$ and the $\mathbb K$-module $V$ possess a common basis $\cal B$, their respective tensor algebras $\otimes_{\mathcal A}({\mathcal M})$ and $\otimes(V)$ possess a common basis made of all monomials in the elements of $\cal B$.
}

\subsection{Vector bundles}

The fibres $\mathbb{V}_m$ of a vector bundle $\mathbb{V}$ over a manifold $M$ are isomorphic to a vector space $V$. The dimension of the fibres $\mathbb{V}_m\cong V$ is called the \textbf{rank of the vector bundle} $\mathbb{V}$.\footnote{The rank of vector bundles will always be assumed constant in this text.} A \textbf{line bundle} is a vector bundle $\mathbb{V}$ of rank one. 

The vector space of sections $\Gamma(\mathbb{V})$ of any vector bundle $\mathbb{V}$ over $M$ is a $ C^\infty(M)$-module, since for any function and any section their pointwise product is also a section: 
\be
f\in  C^\infty(M) \quad\text{and}\quad v\in \Gamma(\mathbb{V})\quad\Longrightarrow\quad  f\cdot v\,\in\,\Gamma(\mathbb{V})\,.
\ee
The vector bundle $\mathbb{V}$ over $M$ is trivial if and only if the $C^\infty(M)$-module $\Gamma(\mathbb{V})$ of its sections is freely-generated (see \textit{e.g.} \cite[Proposition 12.13]{Nestruev}): 
\be
\mathbb{V}=M\times V\quad\Longleftrightarrow\quad\Gamma(\mathbb{V})\,\cong\, C^\infty(M)\otimes V\,.
\ee
Otherwise, the $C^\infty(M)$-module $\Gamma(\mathbb{V})$ is said \textbf{locally free}. Concretely, the generating set is obtained by exhibiting a finite collection of sections $e_i\in \Gamma(\mathbb{V})$ with the property that for each point $m$ the set $\{e_i|_m\}$ spans the fibre $\mathbb{V}_m$. They all are global sections iff the vector bundle is trivial, but they are always guaranteed to exist locally in any trivialisation of the bundle.
A line bundle is trivial iff there exists a nowhere-vanishing global section.
The sections of a trivial bundle $M\times V$, \textit{i.e.} elements of a free $ C^\infty(M)$-module $\Gamma(M\times V)\cong C^\infty(M)\otimes V$, are called $V$-\textbf{valued functions on} $M$.\footnote{Later on, elements of a locally-free $C^\infty(M)$-module will sometimes also be referred to as $V$-valued functions on $M$, with a slight abuse of terminology.} 

The category of trivial vector bundles on a (connected) manifold $M$ is equivalent to the category of free $ C^\infty(M)$-modules.
The Serre-Swan theorem is the generalisation of this result to arbitrary vector bundles.

\subsection{Projective modules over commutative algebras}

There exists a suitable generalisation of the notion of basis. 

Let ${\mathcal B}=\{e_i\,|\,i\in I\}$ be a generating set of the ${\mathcal A}$-module $\cal M$, indexed by the set $I$. A  \textbf{dual basis} is a collection ${\mathcal B}^*=\{e^{*i}\,|\,i\in I\}$ of ${\mathcal A}$-linear maps $e^{*i}:{\mathcal M}\to{\mathcal A}$ such that, for any $v\in\cal M$:
\begin{itemize}
	\item[(i)] $e^{*i}(v)=0$ for almost all $i\in I$ (\textit{i.e.} all except finitely many),
	\item[(ii)] $v=\sum\limits_{i\in I}\,e^{*i}(v)\,e_i$\,.
\end{itemize}
An ${\mathcal A}$-module $\cal M$ is called \textbf{projective} if it admits a dual basis. 
The condition (i) states that the linear combination on the right-hand-side of (ii) always has a finite number of terms.
In particular, the conditions (i)-(ii) imply that any element of the module can be expressed as a finite linear combination of the generating set ${\mathcal B}=\{e_i\,|\,i\in I\}$. One may thus write the following resolution of the identity: 
\be
\sum_{i\in I}\,e_i\otimes e^{*i}=id_{\mathcal M}\,.
\ee
Note that the elements of ${\mathcal B}$ may not be linearly independent and, accordingly, the linear decomposition may not be unique (thus $e^{*i}(e_j)$ may not be equal to the Kronecker delta $\delta^i_j$). 

\vspace{3mm}
\noindent{\small\textbf{Remark:} There exists equivalent (but more abstract) caracterisations of projective modules which are (unfortunately for the layman) more standard in the mathematical literature.
For instance, an ${\mathcal A}$-module $\cal M$ is called \textbf{projective} if any short exact sequence,
\be
0\to {\mathcal K}\stackrel{i}{\hookrightarrow}{\mathcal L}\stackrel{\pi}{\twoheadrightarrow}{\mathcal M}\to 0\,,\label{shortexactM}
\ee
of ${\mathcal A}$-modules, which ends at $\cal M$, splits,
\be
0\leftarrow {\mathcal K}\stackrel{r}{\twoheadleftarrow}{\mathcal L}\stackrel{s}{\hookleftarrow}{\mathcal M}\leftarrow 0\,,\label{splittingM}
\ee
where all arrows are ${\mathcal A}$-module morphisms (\textit{i.e.} ${\mathcal A}$-linear maps).
This abstract definition turns out to be equivalent to the previous one  in terms of the dual basis (see \textit{e.g.} \cite[Proposition 7.58]{Rotman}). It is also equivalent to the following characterisation (see \textit{e.g.} \cite[Theorem 7.56]{Rotman}): an ${\mathcal A}$-module $\cal M$ is projective iff 
there exists another ${\mathcal A}$-module $\cal N$ such that the direct sum ${\mathcal M}\oplus{\mathcal N}$ is a freely-generated ${\mathcal A}$-module.
The terminology ``projective'' may appear elusive at first sight. One possible way to motivate this term is that (due to the previous characterisation) an ${\mathcal A}$-module $\cal M$ is projective iff it is the projection ${\mathcal M}=P{\mathcal F}$ of a free module ${\mathcal F}$ via a projector $P:{\mathcal F}\twoheadrightarrow{\mathcal M}$.
}

\subsection{Serre-Swan theorem}

A finitely-generated projective module is a module admitting a finite dual basis (\textit{i.e.} the indexing set $I$ is finite).
For the present purpose, it will be sufficient to introduce one last bit of terminology which may be more intuitive in the geometric context. Over the structure algebra $C^\infty(M)$ of smooth functions on a manifold $M$, finitely-generated projective modules are sometimes called \textbf{locally-free modules of (locally) finite rank}.\footnote{This terminology will be taken as a mere synonym, without dwelling on subtleties. The algebraic geometer in the room can look at \cite[Theorem 3.4.6]{Raynaud} which provides, under very general assumptions, a criteria for the equivalence.}

A crucial fact is the following proposition: \textit{For any vector bundle $\mathbb{V}$ of finite rank over a manifold $M$, the vector space $\Gamma(\mathbb{V})$ of its global sections is a locally-free $C^\infty(M)$-module of finite rank.} A somewhat surprising corollary is that for any vector bundle $\mathbb{V}$ of finite rank over $M$, there exists a vector bundle $\mathbb{W}$ of finite rank over $M$ such that their direct sum $\mathbb{V}\oplus\mathbb{W}$ is a trivial vector bundle  (see \textit{e.g.} \cite[Proposition 12.33]{Nestruev}).

\vspace{3mm}
\noindent{\small\textbf{Example (Vector fields on flat space):} Consider the Euclidean space $\mathbb{R}^N$ with Cartesian coordinates $X^i$ and basis $\vec e_i$ ($i=1,2,\cdots,N$) so that a position vector $\vec r\in \mathbb{R}^N$ decomposes as $\vec r= X^i\,\vec e_i$\,. Its tangent bundle $T\mathbb{R}^N\cong \mathbb{R}^N\times \mathbb{R}^N$ is a trivial vector bundle of rank $N\in\mathbb N$. Accordingly, the space of vector fields on $\mathbb{R}^N$ is a free $C^\infty(\mathbb{R}^N)$-module $\mathfrak{X}(\mathbb{R}^N)\cong C^\infty(\mathbb{R}^N)\otimes\mathbb{R}^N$ as is manifest in Cartesian coordinates since any vector field $X\in \mathfrak{X}(\mathbb{R}^N)$ reads as $X= X^i(\vec r)\,\partial_i$\,. In the next example, the coordinate basis $\partial_i$ will be identified with the basis $\vec e_i$ with some slight abuse of notation.
}
\vspace{3mm}

\noindent The previous example is somewhat trivial but it allows to construct the following one (inspired from \cite[Example 1]{Swan}).

\vspace{3mm}
\noindent{\small\textbf{Example (Vector fields on the sphere):} Consider the unit sphere $S^n \subset \mathbb{R}^{n+1}$ embedded into the Euclidean space of dimension $N=n+1$, \textit{i.e.} the sphere is realised as the vectors $\vec r\in \mathbb{R}^{n+1}$ such that $\lVert\vec r\rVert=1$. The tangent bundle $TS^n\subset T\mathbb{R}^{n+1}$ is a vector bundle of rank $n$ over the sphere $S^n$. It can be realised as the locus
\be
TS^n\,=\,\{\,(\vec r,\vec v)\in T\mathbb{R}^{n+1}\,:\,\lVert\vec r\rVert=1,\,\, \vec r\cdot\vec v=0\,\}\,,
\ee
where $\cdot$ denotes the scalar product on the Euclidean space $\mathbb{R}^{n+1}$, and it fits into the short exact sequence of vector bundles over $S^n$:
\be\label{sesvectbundles}
0\to TS^n\stackrel{}{\hookrightarrow}S^n\times\mathbb{R}^{n+1}\stackrel{}{\twoheadrightarrow}NS^n\to 0\,,
\ee
where 
\be
S^n\times\mathbb{R}^{n+1}\,=\,\{\,(\vec r,\vec v)\in T\mathbb{R}^{n+1}\,:\,\lVert\vec r\rVert=1,\,\,\vec v\in \mathbb{R}^{n+1}\,\}\,,
\ee
is the restriction of $T\mathbb{R}^{n+1}$ to $S^n$ (in more fancy terms, it is the pullback of $T\mathbb{R}^{n+1}$ along the embedding $S^n\hookrightarrow\mathbb{R}^{n+1}$) and the quotient 
\be
NS^n\,=\,S^n\times\mathbb{R}^{n+1}\,/\,TS^n\,\cong\, S^n\times\mathbb{R}
\ee is the normal bundle. The latter bundle is a trivial line bundle over $S^n$ which can be realised as
\be
NS^n\,=\,\{\,(\vec r,\lambda\,\vec r)\in T\mathbb{R}^{n+1}\,:\,\lVert\vec r\rVert=1,\,\, \lambda\in\mathbb R\,\}\,.
\ee
This embedding $NS^n\subset T\mathbb{R}^{n+1}$ provides a splitting
\be
0\leftarrow TS^n\stackrel{}{\twoheadleftarrow}S^n\times\mathbb{R}^{n+1}\stackrel{}{\hookleftarrow}NS^n\leftarrow 0\,,
\ee
of the short exact sequence \eqref{sesvectbundles}. This shows that $S^n\times\mathbb{R}^{n+1}\cong TS^n\oplus NS^n$, so that $TS^n$ is a summand of a trivial vector bundle.
In terms of the corresponding spaces of global sections, the short exact sequence \eqref{sesvectbundles} of vector bundles over $S^n$ corresponds to the short exact sequence of $C^\infty(S^n)$-modules:
\be
0\to \mathfrak{X}(S^n)\stackrel{i}{\hookrightarrow}C^\infty(S^n)\otimes\mathbb{R}^{n+1}\stackrel{\pi}{\twoheadrightarrow}C^\infty(S^n)\to 0\,,
\ee
where $\pi(\vec v)=\vec v\cdot\vec n$.
This short exact sequence splits and $C^\infty(S^n)\otimes\mathbb{R}^{n+1}\cong C^\infty(S^n)\oplus\mathfrak{X}(S^n)$, The space $\mathfrak{X}(S^n)$ of vector fields on $S^n$ is a summand of a free  $C^\infty(S^n)$-module, hence it is projective $C^\infty(S^n)$-module. It is generated by the finite set 
\be
\mathcal{B}=\{\,\vec \chi_i\in\mathfrak{X}(S^n) \,\mid\, i=1,2,\cdots,n+1\}\,,\quad\text{where}\quad \vec \chi_i\,:=\,\vec e_i\,-\,X_i \, \vec r
\ee
with some slight abuse of notation. Remember that the position vector decomposes as $\vec r= X^i\,\vec e_i$\,. The vector fields $\chi_i$ are clearly not linearly independent since $X^i\vec \chi_i=\vec 0$.
The corresponding dual basis is made of $C^\infty(S^n)$-linear maps $\varphi^i:\mathfrak{X}(S^n)\to C^\infty(S^n)$ given by the projections on the components (in the ambient basis):
\be
\mathcal{B}^*=\{\,\varphi^i\,\mid\, i=1,2,\cdots,n+1\}\,,\quad\text{where}\quad \varphi^i(\vec v)\,:=\,v^i\,=\,\vec e_i\cdot\vec v\,,
\ee
for any $\vec v\,=\,v^i\,\vec e_i\,$.
In fact, one can check that $\vec v\,=\,\varphi^i(\vec v)\,\vec\chi_i$ because $X_i\,\varphi^i(\vec v)=0$ since $\vec r\cdot\vec v=X_i\,v^i=0$.
}
\vspace{3mm}

The idea of the proof (see \textit{e.g.} \cite[Chap.12]{Nestruev}) of the above proposition (the global sections of a vector bundle span a locally-free module of finite rank) is to construct a dual basis of global sections by making use of a  \textbf{partition of unity} of the manifold $M$, \textit{i.e.} a set of nowhere-negative functions $\{f_\alpha\}$ on $M$ such that for every point $m\in M$, there is a neighbourhood of $m$ where all but a finite number of those functions are vanishing, and $\sum_\alpha f_\alpha=1\,$. A partition of unity $\{f_\alpha\}$ is said \textbf{subordinate to an open cover of the manifold} $M$ ($=\bigcup_\alpha U_\alpha$) if the support of $f_\alpha$ is contained inside the open set $U_\alpha$.

\vspace{3mm}
\noindent{\small\textbf{Heuristic of the proof:} Consider a finite collection ${\mathcal B}=\{e_i\,|\,i\in I\}$ (with $|I|<\infty$) of global sections of the vector bundle $\mathbb{V}$ such that the corresponding restrictions ${\mathcal B}_\alpha=\{e_i|_{U_\alpha}\}$ for an open cover $\{U_\alpha\}$ of the base manifold $M$ span the corresponding space of local sections. This means that one can extract a local basis from each collection ${\mathcal B}_\alpha$.
Note that, by assumption, each element $e_i|_{U_\alpha}:U_\alpha\hookrightarrow\mathbb{V}$ is a local section (\textit{i.e.} $\pi\circ e_i|_{U_\alpha}=id_{U_\alpha}$) that extends to a global section, but they need not be linearly independent (even locally). 
 Let us introduce on each neighborhood a local dual basis ${\mathcal B}^*_\alpha=\{e^{*i}_\alpha\}$, \textit{i.e.} $v|_{U_\alpha}=\sum_i e^{*i}_\alpha(v|_{U_\alpha}) e_i|_{U_\alpha}$. Consider a partition of unity $\{f_\alpha\}$ subordinate to the open cover $\{U_\alpha\}\,$.
For any global section $v\in\Gamma(\mathbb{V})$ on has $v=\sum_\alpha f_\alpha \,v=\sum_\alpha f_\alpha v|_{U_\alpha}=\sum_{i,\alpha} e^{*i}_\alpha(v|_{U_\alpha}) f_\alpha e_i$.
Let us define the dual basis elements $e^{*i}:=\sum_\alpha f_\alpha e^{*i}_\alpha\circ |_{U_\alpha}$ where $|_{U_\alpha}$ denotes the restriction map to the neighborhood $U_\alpha$, hence 
$\sum_{i}e^{*i}(v)\,e_i=\sum_{i,\alpha} e^{*i}_\alpha(v|_{U_\alpha})f_\alpha e_i =v$.
The conclusion is that ${\mathcal B}^*=\{e^{*i}\,|\,i\in I\}$ is a dual basis of the ${\mathcal C}^\infty(M)$-module $\Gamma(\mathbb{V})$ and ${\mathcal B}:=\{e_i\,|\,i=1,\cdots,r\}$ is a generating set. \qed
}
\vspace{3mm}

The Serre-Swan representation theorem asserts that the converse is also true\footnote{Strictly speaking, the manifold $M$ was assumed to be an affine variety or a paracompact topological space in the original versions \cite{Swan,Serre} of the theorem but this assumption can be relaxed (see \textit{e.g.} \cite[Chap.12]{Nestruev} for the smooth case or \cite{Morye:2009} for a very broad generalisation).}: every locally-free $C^\infty(M)$-module (of finite rank) arises in this way from some vector bundle on $M$ (of finite rank). There is even a stronger statement, which one quotes here but whose technical details (such as the definition of the equivalence of categories) will not be discussed here (see \textit{e.g.}  \cite[Section 12.33]{Nestruev} for details).

\vspace{3mm}
\noindent{\small\textbf{Serre-Swan representation theorem:} \textit{The category of (smooth) vector bundles of finite rank on a (connected smooth) manifold $M$ is equivalent to the category of locally-free $ C^\infty(M)$-modules of finite rank.}}
\vspace{3mm}

This is another instance of the duality between geometry and algebra where one may equivalently investigate a vector bundle via its geometrical incarnation $\mathbb{V}$ (a fibred manifold) or via its ``dual'' realisation $\Gamma(\mathbb{V})$ (its space of sections).\footnote{This duality is so fundamental in modern geometry that these two dual notions (the vector bundle $\mathbb{V}$ and its vector space $\Gamma(\mathbb{V})$ of global sections) are often denoted by exactly the same symbol in numerous mathematical papers.}  

\vspace{3mm}
\noindent{\small\textbf{Remark:} The section functor $\Gamma$ mapping vector bundles  $\mathbb{V}$ over $M$ to $C^\infty(M)$-modules  $\Gamma(\mathbb{V})$ preserves the tensor product (see \textit{e.g.} \cite[Theorem 12.39]{Nestruev}) in the sense that 
	\be\label{tensprodvectbdles}
	\Gamma(\mathbb{V}\otimes\mathbb{W})\,\cong\,\Gamma(\mathbb{V})\,\otimes_{\,{\mathcal C}^\infty(M)}\,\Gamma(\mathbb{W})\,,
	\ee
where $\mathbb{V}\otimes\mathbb{W}$ denotes the (fibrewise) tensor product bundle of the vector bundles $\mathbb{V}$ and $\mathbb{W}$ (over the same base manifold $M$). The vector bundle associated to the structure algebra $C^\infty(M)$ seen as a module over itself is the trivial line bundle $\mathbb{I}_M:=M\times\mathbb R$ over the manifold $M$, \textit{i.e.} $\Gamma(\mathbb{I}_M)=C^\infty(M)$. It is sometimes called the \textbf{unit bundle} because it is the unit with respect to the tensor product $\mathbb{I}_M\otimes\mathbb{V}\cong \mathbb{V}\cong \mathbb{V}\otimes\mathbb{I}_M$, in agreement with \eqref{tensprodvectbdles}:
	\be
	\Gamma(\mathbb{I}_M\otimes\mathbb{V})\,\cong\,\,C^\infty(M)\otimes_{\,C^\infty(M)}\,\Gamma(\mathbb{V})\,\cong\,\,\Gamma(\mathbb{V})\,.
	\ee
Note that the section functor $\Gamma$ mapping vector bundles  $\mathbb{V}$ over $M$ to $C^\infty(M)$-modules  $\Gamma(\mathbb{V})$ also preserves the direct sum: 	$\Gamma(\mathbb{V}\oplus\mathbb{W})\,\cong\,\Gamma(\mathbb{V})\,\oplus\,\Gamma(\mathbb{W})$ where $\mathbb{V}\oplus\mathbb{W}$ denotes the fibrewise direct sum (aka \textbf{Whitney sum}) of vector bundles over $M$. 
}

\pagebreak

\section{Lie-Rinehart algebras and Lie algebroids}\label{LARalgebras}

In this chapter and the next one, the symbol $\mathcal A$ will always denote a
commutative algebra, whose geometric avatar will be the structure algebra ${\mathcal C}^\infty(M)$ of a manifold $M$.
One of the interest of the geometric notion of Lie algebroid is that it allows to introduce algebraically and thereby unify the notion of flat (or curved) connections on principal and vector bundles. In this way, connections arise naturally as (slightly broken) representations of the infinitesimal symmetries of the base manifold in terms of infinitesimal symmetries of the bundle. This is the point of view which will be advocated here.

\subsection{Lie-Rinehart algebras}\label{LRAlg}

\subsubsection{Definition and examples of Lie-Rinehart algebras}

Let $\mathcal A$ be a commutative algebra.
Let $\mathfrak{L}$ be a vector space equipped with both a Lie algebra structure (\textit{i.e.} a Lie bracket on $\mathfrak{L}$) and a left $\mathcal A$-module structure (\textit{i.e.} an action of $\mathcal A$ on $\mathfrak{L}$). The symbol $\cdot$ will denote both the product in $\mathcal A$ and the left action of $\mathcal A$ on $\mathfrak{L}$.
An \textbf{anchor} is a linear map\,,
\be\label{Anchor}
\rho\,:\,\mathfrak{L} \to \mathfrak{der}({\mathcal A})\,:\,X\mapsto\hat{X}\,,
\ee
which 
\begin{enumerate}
	\item is both an $\mathcal A$-module morphism and a Lie algebra morphism, and	
	\item appears as follows in the generalised Leibniz rule for the Lie bracket
\be
[X, f\cdot\,Y ] \,=\, f\cdot[X,Y] + \hat{X}[f]\cdot Y\,,\qquad \forall {X},{Y}\in \mathfrak{L}\,,\quad\forall f\in{\mathcal A}\,,
\label{LR}
\ee
where $\hat{X}[f]\in\mathcal A$ denotes the image of $f\in\mathcal A$ by the derivation $\rho(X)=\hat{X}\in\mathfrak{der}({\mathcal A})$. Note that by definition $\hat{X}$ obeys the usual Leibnitz rule
\be\label{vectfielder}
\hat{X}[f\cdot g]=\hat{X}[f]\cdot g+f\cdot\hat{X}[g]\,,\quad\forall f,g\in {\mathcal A}\,.
\ee
\end{enumerate}
A \textbf{Lie-Rinehart algebra} over $\mathcal A$ is
a Lie algebra $\mathfrak{L}$, equipped both with a structure of left $\mathcal A$-module and with an anchor.\footnote{As a side historical remark, one can mention that the structure of ``Lie-Rinehart algebras'' was introduced by Herz in the fifties under the name ``Lie pseudo-algebra'' \cite{Herz}. The term ``Lie-Rinehart algebra'' was coined by Huebschmann in the nineties, as a term acknowledging Rinehart's seminal contributions, in particular to the description of its universal enveloping algebra in \cite{Rinehart} (see \cite{Huebschmann,Huebschmann:2021} for more historical remarks).}
In other words, a Lie-Rinehart algebra $\mathfrak{L}$ over $\mathcal A$ is
a pair made of a Lie algebra $\mathfrak{L}$ and a commutative algebra $\mathcal A$, equipped with representations on each other (in fact, $\mathfrak{L}$ is an $\mathcal A$-module via the action $\cdot$ and, conversely, $\mathcal A$ is an $\mathfrak{L}$-module via the anchor $\rho$) compatible with their respective products (in the sense of the corresponding Leibniz rules, \textit{i.e.} \eqref{LR} for the Lie bracket of $\mathfrak{L}$ and \eqref{vectfielder} for the commutative product of $\mathcal A$).

Due to their central importance in this text, a long list of examples of Lie-Rinehart algebras is provided immediately (and more are coming).

\vspace{2mm}\noindent\textbf{Example 1 (Derivations)\,:} The archetypal example of Lie-Rinehart algebra is the Lie algebra $\mathfrak{der}({\mathcal A})$ ($=\mathfrak{L}$) of derivations of $\mathcal A$ endowed with the identity map $id_{\mathfrak{L}}$ ($=\rho$) as anchor.

\vspace{2mm}\noindent\textbf{Example 2 (Vector fields)\,:} As suggested by our choice of notation in \eqref{LR}, a particular instance of Example 1 is the Lie algebra ${\mathfrak{X}}(M)$ ($=\mathfrak{L}$) of vector fields. It forms a Lie-Rinehart algebra over the structure algebra ${\mathcal C}^\infty(M)$ ($={\mathcal A}$).

\vspace{2mm}\noindent\textbf{Example 3 (Lie algebras)\,:} The simplest example of Lie-Rinehart algebra with trivial anchor ($\rho=0$) is a Lie algebra $\mathfrak{g}$ ($=\mathfrak{L}$) over a field $\mathbb K$ ($={\mathcal A}$). More generally, Lie-Rinehart algebras over a commutative algebra ${\mathcal A}$ with trivial anchor are called $\mathcal A$-\textbf{Lie algebras}.

\vspace{2mm}\noindent\textbf{Example 4 (Free modules generated by Lie algebras)\,:} In turn, the simplest class of $\mathcal A$-Lie algebras defined in Example 3 are free ${\mathcal A}$-modules ${\mathcal A}\otimes\mathfrak{g}$ ($=\mathfrak{L}$) generated by a Lie algebra $\mathfrak{g}$. The Lie bracket is simply the ${\mathcal A}$-bilinear extension of the one of $\mathfrak{g}$.

\vspace{2mm}\noindent\textbf{Example 5 (Lie algebra actions)\,:} A representation of a Lie algebra $\mathfrak{g}$ on the commutative algebra ${\mathcal A}$ is a Lie algebra morphism 
\be\label{gaction}
\#\,:\,\mathfrak{g}\to\mathfrak{der}({\mathcal A})\,:\,v\mapsto v^\#\,.
\ee
The free ${\mathcal A}$-module ${\mathcal A}\otimes\mathfrak{g}$ ($=\mathfrak{L}$) is endowed with a Lie-Rinehart algebra structure via the following Lie bracket\footnote{The Jacobi identity is a non-trivial fact if one checks it by brute force computation. A simpler way is to understand that \eqref{actionLiebracket} is the only Lie bracket reducing to the one of $\mathfrak{g}$ and compatible with the Leibnitz rule for the anchor \eqref{anchorLRaction}.}
\ba\label{actionLiebracket}
\llbracket f\otimes v,g\otimes w\rrbracket \,:=\,(fg)\otimes[v,w]\,+\,\big(f\, v^\#(g)\big)\otimes w\,-\,\big(g\, w^\#(f)\big)\otimes v\,,\\
 \forall f,g\in {\mathcal A}\,,\quad\forall v,w\in \mathfrak{g}\,,   \nonumber
\ea
and the ${\mathcal A}$-linear extension 
\be\label{anchorLRaction}
\#\,:\,{\mathcal A}\otimes\mathfrak{g}\to\mathfrak{der}({\mathcal A})\,:\,f\otimes v\mapsto f\cdot v^\#
\ee
of the $\mathfrak{g}$-action \eqref{gaction} as anchor. This Lie-Rinehart algebra is called the \textbf{action (or transformation) Lie-Rinehart algebra} and it will be denoted ${\mathcal A}\rtimes\mathfrak{g}$.
In the particular case of a trivial action ($\#=0$), the corresponding Lie-Rinehart algebra
is a free ${\mathcal A}$-Lie algebra ${\mathcal A}\otimes\mathfrak{g}$ generated by a Lie algebra $\mathfrak{g}$, as in Example 4.

\vspace{2mm}\noindent\textbf{Example 6 (Poisson algebras)\,:} An important (family of) example(s) of Lie-Rinehart algebras are provided by Poisson algebras $\cal P$ which, by definition, can be seen simultaneously as commutative algebras (${\mathcal A}={\mathcal P}$) and Lie algebras ($\mathfrak{L}={\mathcal P}$) which are consequently (bi)modules over themselves. The Poisson bracket gives the Lie bracket and the anchor is defined by the adjoint action of the Poisson bracket ($\rho=ad$). That the latter is indeed an anchor, i.e. it satisfies the Leibniz rules \eqref{LR}-\eqref{vectfielder}, follows from the fact that the Poisson bracket is a biderivation. 

\vspace{2mm}\noindent\textbf{Example 7 (Schouten algebras)\,:} A Poisson algebra ${\mathcal P}$ which is $\mathbb N$-graded as a commutative algebra and such that its suspension ${\mathcal P}[1]$ is $\mathbb N$-graded as a Lie algebra is called a \textbf{Schouten algebra}.\footnote{Note that a Gerstenhaber algebra is the supercommutative analogue of a Schouten algebra.} Concretely, this condition means that the commutative product $\cdot$ of ${\mathcal P}$ is such that ${\mathcal P}_i\cdot{\mathcal P}_j\subseteq {\mathcal P}_{i+j}$ while the Poisson bracket $\{\,,\,\}$ is such that $\{{\mathcal P}_i,{\mathcal P}_j\}\subseteq {\mathcal P}_{i+j-1}$.
In particular, the Lie subalgebra $\mathfrak{L}={\mathcal P}_1\subset{\mathcal P}$ in degree one of any Schouten algebra ${\mathcal P}$, defines a Lie-Rinehart algebra over the commutative subalgebra ${\mathcal A}={\mathcal P}_0\subset{\mathcal P}$ in degree zero, with Poisson bracket as Lie bracket and the adjoint action of ${\mathcal P}_1$ on ${\mathcal P}_0$ as anchor ($\rho=ad|_{{\mathcal P}_1}$).

\subsubsection{Lie-Rinehart subalgebras}

A linear map $F:\mathfrak{L}_1\to\mathfrak{L}_2$ from the Lie-Rinehart algebra $\mathfrak{L}_1$ over $\mathcal A$ to the Lie-Rinehart algebra $\mathfrak{L}_2$ over $\mathcal A$ is a \textbf{morphism between two Lie-Rinehart algebras over the same commutative algebra} if it is both an $\mathcal A$-module morphism (it is a $\mathcal A$-linear) and a Lie algebra morphism (the image of the source Lie bracket is the target Lie bracket of the images), compatible with the two respective anchors $\rho_1$ and $\rho_2$ in the sense that the following diagram commutes
\be
\begin{array}
[c]{ccc}%
\mathfrak{L}_1&\stackrel{F}{\longrightarrow}&\mathfrak{L}_2\\
\rho_1\searrow&&\swarrow\rho_2\\
&\mathfrak{der}({\mathcal A})&
\end{array}
\ee

\vspace{2mm}\noindent{\small\textbf{Example (Lie pairs)\,:}
A Lie-Rinehart subalgebra $\mathfrak{H}\subset\mathfrak{G}$ of a Lie-Rinehart algebra $\mathfrak{G}$ over the same commutative algebra $\mathcal A$ is sometimes called a \textbf{Lie pair}.\footnote{\label{prevfootn}See \textit{e.g.} \cite{Laurent-Gengoux:2014} in the context of Lie algebroids.} The embedding $i:\mathfrak{H}\hookrightarrow\mathfrak{G}$ is an injective morphism between these Lie-Rinehart algebras over $\mathcal A$. Note that the quotient $\mathfrak{G}/\mathfrak{H}$ is an $\mathcal A$-module. Moreover, there is a canonical surjection 
$\pi:\mathfrak{G}\twoheadrightarrow\mathfrak{G}/\mathfrak{H}$.
A Lie pair defines a short exact sequence of $\mathcal A$-modules
\be
0\to \mathfrak{H}\stackrel{i}{
\hookrightarrow}
\mathfrak{G}\stackrel{\pi}{
\twoheadrightarrow
}
\mathfrak{G}/\mathfrak{H}\to 0\,.\label{shortexactinfLie}
\ee
}

\subsection{Lie algebroids}

The geometric avatars of Lie-Rinehart algebras are Lie algebroids. They were introduced by Pradines in the sixities \cite{Pradines:1967}.
 
\subsubsection{Definition and examples of Lie algebroids}

A \textbf{Lie algebroid} over $M$ is a locally-free Lie-Rinehart algebra of finite rank over the structure algebra ${\mathcal C}^\infty(M)$ ($={\mathcal A}$). More concretely, from the Serre-Swan theorem such a Lie-Rinehart algebra arises as the space $\Gamma(\mathbb{A})$ ($=\mathfrak{L}$) of global sections of a vector bundle $\mathbb{A}$ over $M$ with the anchor arising from a morphism
$\rho:\mathbb{A}\to TM$ of vector bundles over $M$ (which is also called the anchor\footnote{The term ``anchor'' was introduced by Mackenzie. According to him, this map ``ties, or fails to tie, the bracket structure'' on the space $\Gamma(\mathbb{A})$ of sections of the vector bundle $\mathbb{A}$ over $M$ to the Lie bracket on the space ${\mathfrak{X}}(M)$ of vector fields on the base manifold \cite[p.101]{Mackenzie}.}).

\vspace{2mm}\textbf{Examples:}
\begin{itemize}
	\item[$\bullet$] \underline{Bi}j\underline{ective anchor:} The tangent bundle $TM$ ($=\mathbb{A}$) of $M$ is the archetype of Lie algebroid over $M$ with bijective anchor ($\rho = $ identity). In fact, a Lie algebroid with bijective anchor is, by definition, isomorphic to the tangent bundle of its base manifold.
	\item[$\bullet$] \underline{Trivial anchor:} Lie algebroids over a single point (so their anchor must vanish identically) are nothing but Lie algebras. Retrospectively, Lie algebras can be thought of as the simplest examples of Lie algebroids. The next-to-simplest examples are the Lie algebroids over a manifold, say $M$, with trivial anchor ($\rho=0$). They are
called ${\mathcal C}^\infty(M)$-\textbf{Lie algebra bundles} (or \textbf{weak Lie algebra bundles}) because each of their typical fibres are endowed with a Lie algebra structure (not necessarily isomorphic to each other). Their space $\Gamma(\mathbb{A})$ of global sections are Lie algebras over ${\mathcal C}^\infty(M)$ with pointwise Lie bracket. Lie algebroids with trivial anchor such that all fibres are isomorphic to a single Lie algebra $\mathfrak{g}$ are sometimes called $\mathbb R$-Lie algebra bundles (or \textbf{strong Lie algebra bundles}) or, most often, simply \textbf{Lie algebra bundles}.\footnote{Lie algebra bundles are integrable, in the sense that the Lie groups integrating each Lie algebra fiber fit into a smooth bundle of Lie groups \cite{Douady}.} Concretely, the structure constants of ${\mathcal C}^\infty(M)$-Lie algebra bundles depend on the base coordinates while the ones of $\mathbb R$-Lie algebra bundles are constant (in a suitable coordinate basis).
	\item[$\bullet$] \underline{In}j\underline{ective anchor:} Another important class of example of Lie algebroids are the ones with injective anchor ($\text{Ker}\,\rho=\{0\}$). They are equivalent to \textbf{involutive distributions} on $M$. Indeed, the image $\rho(\mathbb{A})\subset TM$ defines a distribution on $M$, which is involutive (in the sense that it obeys Frobenius integrability condition) since the Lie bracket of their sections (seen as vector fields on $M$) closes (by the definition of Lie algebroid). The quotient vector bundle $TM/\mathbb{A}$, whose fibres are the quotient of the tangent space to $M$ at each point by the tangent space to the leaf at that point, is called the \textbf{normal bundle} of the foliation. 
	\item[$\bullet$] \underline{Sur}j\underline{ective anchor:} The \textbf{transitive Lie algebroids} are such that their anchor
	is surjective ($\text{Im}\,\rho=TM$).\footnote{Sometimes ${\mathcal C}^\infty(M)$-Lie algebra bundles are called ``totally intransitive'' Lie algebroids, because they are the extreme opposite of transitive Lie algebroids($\text{Im}\,\rho=0$ vs $\text{Im}\,\rho=TM$).} The vector sub-bundle $\text{Ker}\,\rho\subset \mathbb{A}$ of a transitive Lie algebroid is a Lie algebra bundle\footnote{This property is not obvious (see \textit{e.g.} \cite[Theorem IV.1.4]{Mackenzie} for an elegant proof).} called the \textbf{adjoint algebroid} whose typical fibre $\mathfrak{g}$ is called the \textbf{isotropy algebra of the transitive Lie algebroid}. The short exact sequence of Lie algebroid morphisms
\be\label{transitivLiealgebro}
0\to \text{Ker}\,\rho\hookrightarrow \mathbb{A}\stackrel{\rho}{\twoheadrightarrow} TM\to 0
\ee
will be called the \textbf{Atiyah sequence of the transitive Lie algebroid}. 
\end{itemize}

The examples of trivial and injective anchors are more generic than it may appear at first sight. In fact, to any Lie algebroid of anchor 
$\rho:\mathbb{A}\to TM$ one may associate two Lie algebroids on $M$: on the one hand, a ${\mathcal C}^\infty(M)$-Lie algebra bundle $\text{Ker}\,\rho$ whose fibre at $m\in M$ is called the \textbf{isotropy algebra at the point} $m$
and, on the other hand, a (possibly singular) foliation Im$\,\rho$ on $M$ called the \textbf{characteristic foliation}.
The former is a vector sub-bundle of $\mathbb{A}$ while the latter is a vector sub-bundle of $TM$:
\be
\text{Ker}\,\rho\hookrightarrow \mathbb{A}\stackrel{\rho}{\twoheadrightarrow}\text{Im}\,\rho\hookrightarrow TM\,.
\ee

A morphism $\varphi:\mathbb{A}_1\to\mathbb{A}_2$ of vector bundles over $M$ from the Lie algebroid $\mathbb{A}_1$ to the Lie algebroid $\mathbb{A}_2$ is a \textbf{morphism between two Lie algebroids over the same base manifold} $M$ if the corresponding ${\mathcal C}^\infty(M)$-linear map $\Gamma(\varphi)\,:\Gamma(\mathbb{A}_1)\to\Gamma(\mathbb{A}_2)$ between their section spaces is a morphism of Lie-Rinehart algebras over the structure algebra ${\mathcal C}^\infty(M)$. In particular, it must be compatible with the two respective anchors $\rho_1$ and $\rho_2$ in the sense that the following diagram commutes
\be
\begin{array}
[c]{ccc}%
\mathbb{A}_1&\stackrel{\varphi}{\longrightarrow}&\mathbb{A}_2\\
\rho_1\searrow&&\swarrow\rho_2\\
&TM&
\end{array}
\ee

\subsubsection{Infinitesimal actions on a manifold}

A representation of a Lie algebra $\mathfrak{g}$ on the structure algebra ${\mathcal C}^\infty(M)$ of a manifold $M$, \textit{i.e.} a Lie algebra morphism	
\be\label{actionLiealgonM}
\#\,:\,\mathfrak{g}\to {\mathfrak{X}}(M)\,:\,y\mapsto y^\#
\ee
from the Lie algebra $\mathfrak{g}$ to the Lie algebra ${\mathfrak{X}}(M)$ of vector fields on $M$, is called a \textbf{an action of the Lie algebra} $\mathfrak{g}$ \textbf{on the manifold} $M$. Any such action defines a field of linear maps 
\be
\#_m\,:\,\mathfrak{g}\to T_mM\,:\,y\mapsto y^\#|_m\,,
\ee
which in turn leads to a Lie algebroid structure on the trivial vector bundle $M\times\mathfrak{g}$. 
Such a Lie algebroid is called the \textbf{action (or transformation) Lie algebroid} on $M$ associated to the action of $\mathfrak{g}$ on $M$. Accordingly, it is often denoted $M\rtimes\mathfrak{g}$.\footnote{Following \cite{Kosmann-Schwarzbach}, the action Lie algebroid is sometimes denoted $\mathfrak{g}\mathbin{\hbox{$<\kern-.4em\mapstochar\kern.4em$}} M$ to emphasise the action of $\mathfrak{g}$ on $M$. More frequently, the evocative notation $M\rtimes\mathfrak{g}$ is used to stress that, as a vector bundle, the action Lie algebroid is simply the trivial vector bundle $M\times\mathfrak{g}$.} Its anchor reads
\be
\#_\bullet\,:\,M\rtimes\mathfrak{g}\to TM\,:\,(m,y)\mapsto y^\#|_m\,.
\ee
Its space of global sections is the action Lie-Rinehart algebra $\Gamma(M\rtimes \mathfrak{g})\,\cong\,{\mathcal C}^\infty(M)\rtimes\,\mathfrak{g}$ spanned by sections of the trivial vector bundle $M\times\mathfrak{g}$, \textit{i.e.} $\mathfrak{g}$-valued functions on $M$. Its anchor is the ${\mathcal C}^\infty(M)$-linear extension of \eqref{actionLiealgonM}.
The Lie bracket on $\Gamma(M\rtimes \mathfrak{g})\cong {\mathcal C}^\infty(M)\rtimes \mathfrak{g}$ is defined by \eqref{actionLiebracket}, \textit{i.e.} it is the following Lie bracket between $\mathfrak{g}$-valued functions on $M$
\be
\llbracket v,w\rrbracket :=[v,w]_\mathfrak{g}+v^\#[w]-w^\#[v]\,,\qquad \forall v,w\in{\mathcal C}^\infty(M)\otimes \mathfrak{g}\,,
\ee
and is simply the pullback of the Lie bracket of the corresponding vector fields (this makes automatic the Jacobi identity).

\vspace{2mm}
{\small \textbf{Remark:} To make this more explicit, one can introduce a basis  for $\mathfrak{g}$, which will be denoted $\{\texttt{T}_i\}$ with $C^i_{jk}$ the corresponding structure constants: $[\,\texttt{T}_j,\texttt{T}_k]=C^i_{jk}\texttt{T}_i$. One may also introduce local coordinates $x^\mu$ on $M$. The images of the basis elements are vector fields whose expression in components is of the form $\texttt{T}_i^\#=X_i^\mu(x)\partial_\mu$. Their Lie bracket reads $[\,\texttt{T}^\dagger_j,\texttt{T}^\dagger_k]=C^i_{jk}\texttt{T}^\dagger_i$, which becomes the relation $X_j^\mu(x)\partial_\mu X^\nu_k(x)-X_k^\mu(x)\partial_\mu X^\nu_j(x)=C^i_{jk}X_i^\nu(x)$ in components.
The $\mathfrak{g}$-valued functions on $M$ take the form $y(x)=y^i(x)\texttt{T}_i$ and the anchor is simply the map
$y^i(x)\texttt{T}_i\mapsto y^i(x)X_i^\mu(x)\partial_\mu$.
Finally, the Lie bracket
\eqref{actionLiebracket} reads explicitly $\llbracket y(x),z(x)\rrbracket ^i={C}^i_{jk}y^j(x)z^k(x)+y^j(x)X_j^\mu(x)\partial_\mu z^i(x)-z^j(x)X_j^\mu(x)\partial_\mu y^i(x)$.}
\vspace{2mm}

According to the various possible properties of the anchor one has the following cases:
\begin{itemize}
	\item[$\bullet$] \underline{Bi}j\underline{ective anchor:} On one extreme, an action Lie algebroid with bijective anchor is said to define a \textbf{regular action of a Lie algebra}.
	\item[$\bullet$] \underline{Trivial anchor:} On the other extreme, an action Lie algebroid with trivial anchor defines a \textbf{trivial action of a Lie algebra}. The action Lie algebroids with trivial anchor coincide with the trivial Lie algebra bundles (that is to say, trivial as vector bundles).
	\item[$\bullet$] \underline{In}j\underline{ective anchor:} An intermediate situation is the \textbf{free action of a Lie algebra} defined by the property that the corresponding action Lie algebroid has injective anchor. Such an anchor is sometimes called a \textbf{pointwise-faithful  representation} since it defines a field of linear injections $\#_m:\mathfrak{g}\hookrightarrow T_mM:y\mapsto y^\#|_m\,$. The vector fields $y^\#\in \mathfrak{g}^\#$ will be called \textbf{nowhere (or everywhere) vanishing vector fields} on $M$ in the sense that 
\be
\forall\, m\in M\,:\,y^\#|_m\neq0\quad\Leftrightarrow\quad y\neq 0\,,
\ee
or, equivalently, that
\be
\exists\, m\in M\,:\,y^\#|_m=0\quad\Leftrightarrow\quad y= 0\,.
\ee 
The injective anchor defines an involutive distribution on $M$ whose leaves are called the \textbf{orbits of the (locally) free action}.
	\item[$\bullet$] \underline{Sur}j\underline{ective anchor:} Another intermediate situation is the \textbf{transitive action of a Lie algebra} corresponding to the case when the corresponding action Lie algebroid has surjective anchor. 
\end{itemize}
Clearly, by definition an action of a Lie algebra (respectively, group) is regular iff it is both free and transitive.
The action of a Lie group $G$ on a manifold $M$ is called a \textbf{locally free action} of a Lie group if all stabiliser subgroups are discrete.
A $G$-action is locally free iff the corresponding $\mathfrak{g}$-action is free. If the
$G$-action is transitive then so is the $\mathfrak{g}$-action. The converse holds if $M$ is connected. 

\subsubsection{Atiyah algebroids}

Although we will neither spell out nor use the notion of Lie groupoids here, let us mention that Lie algebroids are to Lie groupoids what Lie algebras are to Lie groups. For instance, the tangent bundle $TM$ over $M$ is the Lie algebroid of the pair groupoid $M\times M$. Similarly, a theorem of Dazord ensures that any action Lie algebroid $M\rtimes \mathfrak{g}$ is integrable \cite{Dazord}. In practice, this means that there always exists a corresponding action on the manifold $M$ of a Lie group $G$ whose Lie algebra is $\mathfrak{g}$. However, note that the so-called ``Lie's third theorem'' (stating that every finite-dimensional Lie algebra $\mathfrak g$ is associated to a Lie group $G$) does not hold in general for finite-rank Lie algebroids (see \textit{e.g.} the lecture notes \cite{Crainic} on this integrability problem).
For instance, not all transitive Lie algebroids are integrable to Atiyah algebroids.

\vspace{3mm}
\noindent{\small\textbf{Example (Atiyah algebroids)}\,: 
Let $P$ be the total space of a principal $H$-bundle over $P/H$; equivalently, the Lie group $H$ acts freely (and properly) on the total space $P$ 
 and the base $P/H$ is the space of orbits. The transformation Lie algebroid $P\rtimes\mathfrak{h}$ for the free action of $H$ on $P$ is endowed with a fibrewise injective anchor.
The Lie group $H$ also acts freely on the tangent bundle $TP$ of the total space, thus one may consider the space of orbits $TP/H$, which is a Lie algebroid over $P/H$, called the \textbf{Atiyah algebroid of the principal} $H$-\textbf{bundle} $P$. Its global sections are the $H$-invariant vector fields on $P$, which span the Atiyah algebra $\Gamma(TP/H)=\mathfrak{X}(P)^H$. These vector fields can be interpreted as infinitesimal automorphisms of the principal bundle since they generate diffeomorphisms of $P$ which are equivariant with respect to the $H$-action.\footnote{Since vertical automorphisms of a principal bundle are what physicists would call a gauge transformation in Yang-Mills theory (or a local Lorentz transformation in Einstein-Cartan gravity), vertical invariant vector fields can be thought as infinitesimal gauge transformations. This explains our slogan in the introduction.}
The vertical distribution $VP\subset TP$ is canonically isomorphic, as a Lie algebroid over the total space $P$, to the transformation Lie algebroid: $VP\simeq P\rtimes\mathfrak{h}$.
The Lie group $H$ also acts freely on the vertical distribution $VP$ of the total space; thus, one may consider the space of orbits $VP/H$, which is canonically isomorphic to the adjoint algebroid $P\times_{ad_H} \mathfrak{h}$.
Any Atiyah algebroid of a principal $H$-bundle $P$ is a transitive Lie algebroid, the Atiyah sequence of which
is the following short exact sequence \cite{Atiy}
\begin{equation}
\label{Asequ}
0 \to \frac{VP}{H} \hookrightarrow \frac{TP}{H} \twoheadrightarrow T\frac{P}{H} \to 0   
.\end{equation}
of Lie algebroids over $P/H$.}
\vspace{3mm}

A transitive Lie algebroid $\mathbb A$ over a manifold $M$ is integrable (to an Atiyah groupoid) iff it is the Atiyah algebroid of a principal $H$-bundle $P$, \textit{i.e.} ${\mathbb A}=\frac{TP}{H}$ and $M=\frac{P}{H}$\,.
In such case, the Atiyah sequence \eqref{transitivLiealgebro} of the transitive Lie algebroid coincides with the Atiyah sequence \eqref{Asequ} of the principal bundle. In this sense, the algebraic structure of transitive Lie algebroid is the infinitesimal data of principal bundles (like the structure of Lie algebra is the infinitesimal data of Lie groups). Note, however, that not all transitive Lie algebroids are integrable \cite{Almeida}, although they infinitesimally 
look alike.

\subsection{Differential operators acting on modules}\label{diffopsonmodules}

In order to provide the algebraic definition of covariant derivatives and connections in terms of Lie algebroids as well as discussing universal enveloping algebras of the latter, some material on differential operators on vector bundles must be introduced.
 
\subsubsection{Differential operators on modules}

Let $\mathcal A$ be a commutative algebra and let $\textsc{V}$ be an $\mathcal A$-module. For simplicity, the symbol $\cdot$ will denote both the product in $\mathcal A$ and the left action of $\mathcal A$ on $\textsc{V}$. 

The associative algebra of $\mathcal A$-module endomorphisms, \textit{i.e.} of $\mathcal A$-linear maps from $\textsc{V}$ to $\textsc{V}$, will be denoted $\text{End}_{\mathcal A}(\textsc{V})$. The ${\mathcal A}$-linear map $\hat{f}\in\text{End}_{\mathcal A}(\textsc{V})$ acting on $\textsc{V}$ by multiplication by the element $f\in\mathcal A$, \textit{i.e.} 
\be\label{hatf}
\hat{f}\,:\,\textsc{V}\to\textsc{V}\,:\,v\mapsto f\cdot v\,,
\ee 
will be called a \textbf{scalar endomorphism}. The associative algebra of all scalar endomorphisms of the $\mathcal A$-module $\textsc{V}$ is a commutative subalgebra of the associative algebra $\text{End}_{\mathcal A}(\textsc{V})$ of $\mathcal A$-module endomorphisms. It will be denoted
$\hat{\mathcal A}(\textsc{V})$.
There is a canonical embedding of algebras
\be\label{hatcanonic}
\hat{\bullet}\,:\,{\mathcal A}\hookrightarrow \text{End}_{\mathcal A}(\textsc{V})\,:\,f\mapsto\hat{f}\,,
\ee
whose corestriction $\hat{\bullet}:{\mathcal A}\stackrel{\sim}{\to}\hat{\mathcal A}(\textsc{V})$ is an algebra isomorphism between the commutative algebra ${\mathcal A}$ and the algebra $\hat{\mathcal A}(\textsc{V})$ of scalar endomorphisms of the $\mathcal A$-module $\textsc{V}$. The existence of this embedding simply reflects that any element $f\in{\mathcal A}$ of the commutative algebra can be interpreted as a scalar endomorphism $\hat{f}\in\text{End}_{\mathcal A}(\textsc{V})$ via the left action.

Following Grothendieck \cite[Section 16.8]{Grothendieck:1967}, a linear operator $\hat{D}:\textsc{V}\to\textsc{V}$ acting on an ${\mathcal A}$-module $\textsc{V}$ such that, for any set $\{f_1, f_2, \ldots,f_k\}\subset{\mathcal A}$ of $k$ elements in the commutative algebra ${\mathcal A}$, the corresponding $k$-fold nested commutator is $\mathcal A$-linear, \textit{i.e.}
\be
[\,\ldots\,[\,[\hat{D}\,,\,\hat{f}_1]\,,\,\hat{f}_2]\ldots \,,\,\hat{f}_k]\in \text{End}_{\mathcal A}(\textsc{V})\,,
\ee
is called a (linear) \textbf{differential operator of order} $k$ \textbf{acting on the} ${\mathcal A}$-\textbf{module} $\textsc{V}$. The filtered associative algebra of such differential operators will be denoted ${\mathcal D}_{\mathcal A}(\textsc{V})$.
If the module identifies with the commutative algebra, $\textsc{V}={\mathcal A}$, then one gets the \textbf{Grothendieck algebra} ${\mathcal D}({\mathcal A})$ of differential operators acting on ${\mathcal A}$.

\vspace{3mm}
\noindent{\small\textbf{Example (Differential operators on a vector bundle)\,:} 
Consider the particular case where the commutative algebra ${\mathcal A}={\mathcal C}^\infty(M)$ is the structure algebra of a manifold $M$ and the module is locally-free and of finite rank, hence $\textsc{V}=\Gamma(\mathbb{V})$ is the section space of a vector bundle $\mathbb{V}$ over $M$. The corresponding differential operators $\hat{D}:\Gamma(\mathbb{V})\to\Gamma(\mathbb{V})$ are called \textbf{differential operators on the vector bundle} $\mathbb V$.}
\vspace{3mm}

There exist an equivalent definition of the filtered associative algebra ${\mathcal D}_{\mathcal A}(\textsc{V})$ of differential operators (by recursion). 
Let ${\mathcal F}\subset \text{End}_{\mathbb{R}}(\textsc{V})$ be a filtered subalgebra of the associative algebra of endomorphisms of the vector space $\textsc{V}$ such that in degree zero it coincides with the commutative algebra of ${\mathcal A}$-linear maps, ${\mathcal F}_0=\text{End}_{\mathcal A}(\textsc{V})$. A linear operator $\hat{D}\in{\mathcal F}_k$ will be called \textbf{almost} $\mathcal A$-linear if its commutator with any $\mathcal A$-linear map is of lower degree, \textit{i.e.} $[\hat{D},\hat{f}]\in{\mathcal F}_{k-1}$ for any $\hat{f}\in{\mathcal F}_0$. This is just a fancy way to say that a differential operator of order $k$ is $\mathcal A$-linear up to a differential operator of order $k-1$ or lower.
The filtered associative algebra ${\mathcal D}_{\mathcal A}(\textsc{V})$ of differential operators acting on the ${\mathcal A}$-module $\textsc{V}$
is the maximal filtered space of such almost $\mathcal A$-linear operators. 

\subsubsection{Almost-commutative algebras and Lie-Rinehart algebras}

Any associative algebra $\mathcal U$ can be endowed with a Lie algebra structure
via the commutator as Lie bracket, in which case it will be called the \textbf{commutator algebra} and denoted $\mathfrak{U}$ in order to distinguish it from its associative counterpart. 

An \textbf{almost-commutative algebra} $\cal U$ is an $\mathbb N$-filtered associative algebra whose filtration is such that the associated $\mathbb N$-graded algebra $\text{gr}\,\cal U$ is a Schouten algebra (\textit{i.e.} ${\mathcal U}_i\,{\mathcal U}_j\subseteq {\mathcal U}_{i+j}$ and $[{\mathcal U}_i,{\mathcal U}_j]\subseteq {\mathcal U}_{i+j-1}$).
The latter Schouten algebra $\text{gr}\,\cal U$ is sometimes called the \textbf{classical limit of the almost-commutative algebra} $\cal U$. 

The example 7 of Section \ref{LRAlg} implies that the Lie subalgebra $\mathfrak{U}_1\subset\mathfrak{U}$ in degree one of the commutator algebra of any almost-commutative algebra ${\mathcal U}$ defines a Lie-Rinehart algebra over the commutative subalgebra ${\mathcal U}_0\subset{\mathcal U}_1$ in degree zero.
More precisely, the Lie-Rinehart algebra $\mathfrak{U}_1$ over ${\mathcal U}_0$ is an extension of the 
Lie-Rinehart algebra $\mathfrak{U}_1/\mathfrak{U}_0$ (\textit{i.e.} the grade one component of the classical limit algebra $\text{gr}\,\mathfrak{U}$) by the Abelian Lie-Rinehart ideal $\mathfrak{U}_0\subset\mathfrak{U}_1$ in degree zero, in the sense that
\be\label{U0U1quotient}
0\to \mathfrak{U}_0\stackrel{i}{\hookrightarrow}\mathfrak{U}_1\stackrel{\pi}{\twoheadrightarrow}\mathfrak{U}_1/\mathfrak{U}_0\to 0
\ee
is a short exact sequence of Lie-Rinehart algebras over ${\mathcal U}_0$.
The quotient $\mathfrak{U}_1/\mathfrak{U}_0$ is indeed a Lie-Rinehart algebra over ${\mathcal U}_0$, endowed with a Lie bracket and an anchor via the commutator bracket (in the sense that the adjoint representation of $\mathfrak{U}_1$ on 
the subalgebra $\mathfrak{U}_0$ defines the anchor) and with a left ${\mathcal U}_0$-module structure via left multiplication.

\vspace{3mm}
\noindent{\small\textbf{Example (First-order differential operators)\,:} 
The differential operators on the commutative algebra $\mathcal A$ span the Grothendieck algebra ${\mathcal D}({\mathcal A})$ which is almost-commutative.
Its commutator algebra is a filtered Lie algebra which will be denoted $\mathfrak{D}({\mathcal A})$. 
The associative subalgebra of zeroth-order differential operator is isomorphic to the commutative algebra ${\mathcal D}^0({\mathcal A})\cong\mathcal A$.
The Lie subalgebras $\mathfrak{der}({\mathcal A})\subset\mathfrak{D}({\mathcal A})$ and $\mathfrak{D}^1({\mathcal A})\subset\mathfrak{D}({\mathcal A})$ of, respectively, derivations and first-order differential operators are Lie-Rinehart algebra over $\mathcal A$. Moreover, the latter decomposes as the semidirect sum of the former algebras: 
\be
\mathfrak{D}^1({\mathcal A})\cong\mathfrak{der}({\mathcal A})\inplus\mathfrak{D}^0({\mathcal A})
\ee
}

\noindent{\small\textbf{Counter-example (Differential operators on a module)\,:} 
The differential operators on an $\mathcal A$-module $\textsc V$ form a filtered associative algebra ${\mathcal D}_{\mathcal A}(\textsc{V})$ which is \textit{not} almost-commutative in general (when the rank of $V$ is greater than one). In particular, the associative subalgebra ${\mathcal D}^0_{\mathcal A}(\textsc{V})=\text{End}_{\mathcal A}(\textsc{V})$ of zeroth-order differential operators is \textit{not} commutative in general. The 
associative algebra ${\mathcal D}_{\mathcal A}(\textsc{V})$ and its commutator algebra $\mathfrak{D}_{\mathcal A}(\textsc{V})$ are filtered (associative vs Lie, respectively) algebras sharing the same filtration. Therefore, the subspace $\mathfrak{D}^1_{\mathcal A}(\textsc{V})$ of first-order differential operators on an $\mathcal A$-module $\textsc V$ is \textit{not} a Lie subalgebra in general.
}

\subsubsection{Differential operators between modules}

Let $\mathcal A$ be a commutative algebra. Let $\textsc{V}$ and $\textsc{W}$ be two $\mathcal A$-modules.
For simplicity, the canonical embeddings of the commutative algebra ${\mathcal A}$ inside the associative algebras $\text{End}_{\mathcal A}(\textsc{V})$ and $\text{End}_{\mathcal A}(\textsc{W})$ of $\mathcal A$-module endomorphisms will be denoted by the same symbol: one will write ${\mathcal A}\subset\text{End}_{\mathcal A}(\textsc{V})$ and ${\mathcal A}\subset\text{End}_{\mathcal A}(\textsc{W})$ with slight abuse of notation.

A linear operator $\hat{D}:\textsc{V}\to\textsc{W}$ mapping the ${\mathcal A}$-module $\textsc{V}$ to the ${\mathcal A}$-module $\textsc{W}$ such that, for any set $\{f_1, f_2, \ldots,f_k\}\subset{\mathcal A}$ of $k$ elements in the commutative algebra ${\mathcal A}$, the $k$th commutator with each of them is $\mathcal A$-linear, \textit{i.e.}
\be\label{kthcommutatorX}
[\,\ldots\,[\,[\hat{D}\,,\,\hat{f}_1]\,,\,\hat{f}_2]\ldots \,,\,\hat{f}_k]\in \text{Hom}_{\mathcal A}(\textsc{V},\textsc{W})\,,
\ee
is called a (linear) $\textsc{W}$\textbf{-valued differential operator of order} $k$ \textbf{acting on the} ${\mathcal A}$-\textbf{module} $\textsc{V}$.
If the source and target modules are distinct (\textit{i.e.} $\textsc{V}\neq\textsc{W}$) then the $\textsc{W}$-valued differential operator of order $k$ acting on $\textsc{V}$ do \textit{not} form an associative algebra (since they cannot be composed). Nevertheless, they span a filtered $\mathcal A$-bimodule, which will be denoted ${\mathcal D}_{\mathcal A}(\textsc{V},\textsc{W})$.

\vspace{3mm}
\noindent{\small\textbf{Example (Differential operators between vector bundles)\,:} 
Consider the particular case where the commutative algebra ${\mathcal A}={\mathcal C}^\infty(M)$ is the structure algebra of a manifold $M$ and the two modules are locally-free and of finite rank, hence $\textsc{V}=\Gamma(\mathbb{V})$ and $\textsc{W}=\Gamma(\mathbb{W})$ are the section spaces of two vector bundles $\mathbb{V}$ and $\mathbb{W}$ over $M$ with typical fibres the two vector spaces $V$ and $W$. The corresponding differential operators $\hat{D}:\Gamma(\mathbb{V})\to\Gamma(\mathbb{W})$ are called $W$-\textbf{valued differential operators on the vector bundle} $\mathbb V$.}

\subsection{Extensions of Lie-Rinehart algebras}

A Lie-Rinehart subalgebra $\mathfrak{I}\subset\mathfrak{G}$ of a Lie-Rinehart algebra $\mathfrak{G}$ which is both a Lie ideal of $\mathfrak{G}$ (\textit{i.e.} $[\mathfrak{G},\mathfrak{I}]\subset\mathfrak{I}$) and an $\mathcal A$-Lie algebra (\textit{i.e.} it has trivial anchor $\rho|_{{}_\mathfrak{I}}=0$) is called a \textbf{Lie-Rinehart ideal} of the Lie-Rinehart algebra $\mathfrak{G}$\,. The Leibnitz rule \eqref{LR} indeed implies that the anchor of $\mathfrak{I}$ must be trivial in order to be compatible with the condition that it is a Lie ideal.

\vspace{3mm}
\noindent{\small\textbf{Example (Kernel)\,:} The image of the kernel of a morphism $F:\mathfrak{I}\to\mathfrak{G}$ of Lie-Rinehart algebras is a Lie-Rinehart ideal $F(\mathfrak{I})\subset\mathfrak{G}$ of the Lie-Rinehart algebra $\mathfrak{G}$.}
\vspace{3mm}

A short exact sequence of morphism of Lie-Rinehart algebras over the same commutative algebra $\mathcal A$,
\be
0\to \mathfrak{I}\stackrel{i}{\hookrightarrow}\mathfrak{G}\stackrel{\pi}{\twoheadrightarrow}\mathfrak{H}\to 0\,,
\label{shortexactLRalg}
\ee
is called a \textbf{Lie-Rinehart algebra extension} $\mathfrak{G}$ \textbf{of} $\mathfrak{H}$ \textbf{by} $\mathfrak{I}$ \cite{huebschmann1997extensions}.
The short exact sequence \eqref{shortexactLRalg} expresses that $\mathfrak{I}$ is a Lie-Rinehart ideal of $\mathfrak{G}$ and that the corresponding quotient $\mathfrak{G}/\mathfrak{I}$ is isomorphic to the Lie-Rinehart algebra $\mathfrak{H}$. The extensions in the category of finite-rank locally-free Lie-Rinehart algebras over the structure algebra ${\mathcal C}^\infty(M)$ of a manifold $M$ will be called \textbf{Lie algebroid extensions}. 

\vspace{3mm}
\noindent{\small\textbf{Example (Lie-Rinehart algebra $\mathfrak{L}$ with surjective anchor)\,:} 
Any Lie-Rinehart algebra $\mathfrak{L}$ ($=\mathfrak{G}$) with surjective anchor $\rho$ ($=\pi$) defines a Lie-Rinehart algebra extension of the Lie-Rinehart algebra $\mathfrak{der}({\mathcal A})$ of derivations by the $\mathcal A$-Lie algebra $\text{Ker}\,\rho$ ($=\mathfrak{I}$). Conversely, any Lie-Rinehart algebra extension of the Lie-Rinehart algebra of derivations by an $\mathcal A$-Lie algebra is a Lie-Rinehart algebra with surjective anchor.}

\vspace{3mm}
\noindent{\small\textbf{Example (Transitive Lie algebroid)\,:} The Atiyah sequence \eqref{transitivLiealgebro} of any transitive Lie algebroid defines a Lie algebroid extension $\mathbb{A}$ of the tangent bundle $TM$ by the adjoint algebroid $\text{Ker}\,\rho$. Conversely, any Lie algebroid extension of the tangent bundle by a Lie algebra bundle is a transitive Lie algebroid.}

\vspace{3mm}
\noindent{\small\textbf{Example (First-order differential operators on a commutative algebra)\,:} 
One may consider the following short exact sequence
of morphisms of Lie-Rinehart algebras over $\mathcal A$:
\be\label{sxsD1A}
0\to{\mathfrak{D}}^0({\mathcal A})\stackrel{i}{\hookrightarrow}{\mathfrak{D}}^1({\mathcal A})\stackrel{\sigma}{\twoheadrightarrow}\mathfrak{der}({\mathcal A})\to 0\,.
\ee
In other words, the Lie-Rinehart algebra $\mathfrak{D}^1({\mathcal A})$ of first-order differential operators on the commutative algebra $\mathcal A$ is an extension of the Lie-Rinehart algebra $\mathfrak{der}({\mathcal A})$ of derivations of $\mathcal A$ by the Abelian $\mathcal A$-Lie algebra $\mathfrak{D}^0({\mathcal A})\cong\mathcal A$ spanned by zeroth-order differential operators on $\mathcal A$.
}

\subsection{Semidirect sums as split extensions of Lie-Rinehart algebras}

An $\mathcal A$-linear splitting 
\be
0\leftarrow \mathfrak{I}\stackrel{r}{\twoheadleftarrow}\mathfrak{G}\stackrel{s}{\hookleftarrow}\mathfrak{H}\leftarrow 0\,,
\label{splittingLRalg}
\ee
of the short exact sequence \eqref{shortexactLRalg} of $\mathcal A$-module morphisms 
will be called a \textbf{linearly-split extension} $\mathfrak{G}$ \textbf{of} $\mathfrak{H}$ \textbf{by} $\mathfrak{I}$. If the section $s:\mathfrak{H}\hookrightarrow\mathfrak{G}$ is a morphism of Lie-Rinehart algebras, then it
is called a \textbf{Lie-Rinehart split extension} $\mathfrak{G}$ \textbf{of} $\mathfrak{H}$ \textbf{by} $\mathfrak{I}$ \cite{huebschmann1997extensions}). In such case, one will write $\mathfrak{G}=\mathfrak{H}\inplus\mathfrak{I}$, which will be called a \textbf{decomposition as a semidirect sum of Lie-Rinehart algebras}.\footnote{Another definition of a semidirect sum of Lie-Rinehart algebras will be given in Subsection \ref{semidirectsumsLR} in terms of Lie-Rinehart algebra representations.} The split extensions and semidirect sums of Lie algebroids are defined analogously in terms of split extensions and semidirect sums of their locally-free Lie-Rinehart algebras over the structure algebra ${\mathcal C}^\infty(M)$ of their base manifold $M$.

\vspace{3mm}
\noindent{\small\textbf{Example (Direct sums as trivial extensions)\,:} If all arrows of the splitting \eqref{splittingLRalg} are morphisms of Lie-Rinehart algebras, then necessarily the Lie bracket between $\mathfrak{I}$ and $\mathfrak{H}$ vanishes identically. This simplest case of Lie-Rinehart split extension is called a \textbf{trivial extension} and, clearly, $\mathfrak{G}=\mathfrak{H}\oplus\mathfrak{I}$ is the direct sum of the two other Lie-Rinehart algebras. Equivalently, this situation is characterised by the fact that $\mathfrak{G}$ is simulatenously a Lie-Rinehart algebra extension of $\mathfrak{H}$ by $\mathfrak{I}$ and of $\mathfrak{I}$ by $\mathfrak{H}$.
}

\vspace{3mm}
\noindent{\small\textbf{Example (First-order differential operators on a commutative algebra)\,:} 
There is a canonical splitting of $\mathcal A$-module morphisms
\be\label{sxsD1Asplitting}
0\leftarrow{\mathfrak{D}}^0({\mathcal A})\,\twoheadleftarrow\,{\mathfrak{D}}^1({\mathcal A})\,\hookleftarrow\,\mathfrak{der}({\mathcal A})\leftarrow 0\,.
\ee
of the short exact sequence \eqref{sxsD1A}
of Lie-Rinehart algebras, either via the retraction 
\be
\bullet[1]\,:\,{\mathcal D}^1({\mathcal A})\twoheadrightarrow{\mathcal A}\,:\,\hat{D}\mapsto\hat{D}[1]
\ee
given by the evaluation on the unit element $1\in\mathcal A$ (composed with the canonical isomorphism from ${\mathcal A}$ to ${\mathcal D}^0({\mathcal A})$\,) or, equivalently, via the canonical embedding $\mathfrak{der}({\mathcal A})\subset{\mathfrak{D}}^1({\mathcal A})$ (where derivations are seen as particular cases of first-order differential operators).
Note that the latter embedding, \textit{i.e.} the injective arrow in \eqref{sxsD1Asplitting}, is a morphism of Lie-Rinehart algebras over $\mathcal A$.
In other words, the Lie-Rinehart algebra of first-order differential operators on a commutative algebra is the canonical split extension of the Lie-Rinehart algebra of derivations by the Abelian ideal of zeroth-order differential operators.
}

\subsection{Automorphisms of modules}

The aim of this section is to motivate covariant derivatives as infinitesimal symmetries of vector bundles and linear connections as geometric devices lifting base vector fields to such infinitesimal symmetries.

\subsubsection{Finite automorphisms of modules}

Let $\textsc{V}$ and $\textsc{W}$ be two modules over the respective algebras $\mathcal A$ and $\cal B$ with $\cdot$ denoting both the action and product of $\mathcal A$ and with $\bullet$ denoting both the action and product of $\cal B$. 

A morphism $F:{\mathcal A}\to{\mathcal B}$ of algebras defines an $\mathcal A$-module structure on the $\cal B$-module $\textsc W$ via the action of any element $a\in{\mathcal A}$
on any vector $w\in\textsc{W}$ defined as 
\be
a\cdot w:=F(a)\bullet w\,.
\ee
The corresponding $\mathcal A$-module is denoted $F^*\textsc{W}$ and the action of $\mathcal A$ on $\textsc{W}$ is called the \textbf{restriction of scalars}.

A pair made of a morphism $F:{\mathcal A}\to{\mathcal B}$ of algebras, \textit{i.e.} a $\mathbb K$-linear map such that $F(a\cdot b)=F(a)\bullet F(b)$, and a morphism $\varphi:\textsc{V}\to F^*\textsc{W}$ of $\mathcal A$-modules, \textit{i.e.} an $\mathcal A$-linear map $\varphi(a\cdot v)=a\cdot\varphi(v)$, may be thought of as a sort of ``generalised morphism'' from the $\mathcal A$-module $\textsc V$ to the $\cal B$-module $\textsc W$. 
In practice, $\varphi$ is a $\mathbb K$-linear map from the $\mathcal A$-module $\textsc V$ to the $\cal B$-module $\textsc W$ which is $\mathcal A$-\textbf{semilinear} in the sense that 
\be\label{autcond}
\varphi(a\cdot v)=F(a)\bullet\varphi(v)\,, \quad \forall a\in{\mathcal A}\,,\,\,\forall v\in \textsc{V}\,.
\ee
If $F:\mathcal{A}\stackrel{\sim}{\to} \mathcal{A}$ is an automorphism of the commutative algebra $\mathcal{A}$ and $\varphi:\textsc{V}\stackrel{\sim}{\to}F^*\textsc{V}$ is an isomorphism between the $\mathcal{A}$-module $\textsc{V}$ and the $\mathcal{A}$-module $F^*\textsc{V}$, then this pair of morphisms will be called a \textbf{semilinear automorphism} of the $\mathcal{A}$-module $\textsc V$.

\vspace{3mm}
\noindent{\small\textbf{Example (Vector bundle automorphisms)\,:} By the Serre-Swan theorem, if $\textsc{V}$ is a finite-rank locally-free ${\mathcal C}^\infty(M)$-module, then $\textsc{V}$ identifies with the section space $\Gamma(\mathbb{V})$ of a vector bundle $\mathbb{V}$ over $M$. The semilinear automorphisms of the ${\mathcal C}^\infty(M)$-module $\textsc{V}=\Gamma(\mathbb{V})$ are the algebraic versions of the automorphisms of the vector bundle $\mathbb{V}$, \textit{i.e.} diffeomorphisms of $\mathbb{V}$ that linearly map fibres to fibres (see \textit{e.g.} \cite[Sect.4]{Kosmann-Schwarzbach}). On the one hand, the automorphism $F$ of the structure algebra ${\mathcal C}^\infty(M)$ corresponds to a diffeomorphism of the base $M$. On the other hand, the isomorphism $\varphi$ between the ${\mathcal C}^\infty(M)$-module $\Gamma(\mathbb{V})$ and the corresponding restriction of scalars, corresponds to an automorphism of the vector bundle $\mathbb{V}$. If the automorphism of the structure algebra of functions reduces to the identity, \textit{i.e.} $F=id_{{\mathcal C}^\infty(M)}$, then the isomorphism is an automorphism of the ${\mathcal C}^\infty(M)$-module $\Gamma(\mathbb{V})$ corresponding to a vertical automorphism of the vector bundle $\mathbb{V}$, \textit{i.e.} a diffeomorphism of $\mathbb{V}$ linearly mapping each fibre into itself.
}

\subsubsection{Infinitesimal automorphisms of modules}

An \textbf{infinitesimal automorphism} of an $\mathcal A$-module $\textsc V$ is a pair made of a derivation ${\hat X}\in\mathfrak{der}({\mathcal A})$ of the algebra $\mathcal A$ and a vector space endomorphism $\hat{\nabla}\in\mathfrak{gl}(\textsc{V})$, obeying to the Leibnitz rule
\be\label{forallsigma}
\hat{\nabla}(f\cdot \sigma)={\hat X}[f]\cdot \sigma+f\cdot \hat{\nabla}\sigma\,, \quad \forall f\in{\mathcal A}\,,\,\,\forall\sigma\in V\,.
\ee
which is the infinitesimal version of 
\be
\varphi_t(f\cdot \sigma)=F_t(f)\cdot\varphi_t(\sigma)\,, \qquad\forall f\in{\mathcal A}\,,\,\,\forall\sigma\in V\,,
\ee
where $F_t=id_{\mathcal A}+t\,{\hat X}+{\mathcal O}(t^2)$ and $\varphi_t=id_{\textsc V}+t\,{\hat\nabla}+{\mathcal O}(t^2)$ define a one-parameter group of semilinear automorphisms of the $\mathcal A$-module $\textsc V$.
This is the  one-parameter group version of \eqref{autcond}. One used suitable notations for the infinitesimal elements in order to anticipate their later interpretation in the context of linear connections.
The endomorphism $\hat{\nabla}$ will be called a \textbf{covariant derivative} on the $\mathcal A$-module $\textsc V$ \textbf{along the derivation} ${\hat X}$.
\footnote{A side historical remark is that the algebraic properties of covariant derivatives were already stated explicitly by Schouten in 1924 in his book \cite{Schouten}. One should stress that, despite the old history of this algebro-geometric notion, there is no generally accepted terminology for such objects in the mathematical physics literature. For instance, over the years they have been refered to as (the list is not exhaustive) successively as: ``pseudo-linear endomorphisms'' \cite{Jacobson:1935},   ``quasi-scalar differential operators'' \cite{Palais:1968}, ''derivative endomorphisms'' \cite{Kosmann:1976}, ``covariant differential operators'' \cite{Mackenzie}, ... They were also called derivative of $\textsc V$ over the derivation ${\hat X}$ \cite{Kosmann-Schwarzbach}. Essentially, the latter two terminologies of Mackenzie of Kosmann-Schwarzbach are nicely combined in the standard colloquial term ``covariant derivative'' which may be the more evocative for theoretical physicists and closer to its informal use in their daily work.}

Any covariant derivative is a first-order differential operator acting on the $\mathcal A$-module $\textsc V$, \textit{i.e.} $\hat{\nabla}\in{\mathcal D}_{\mathcal A}^1(\textsc{V})$, since the equality \eqref{forallsigma} for all $\sigma\in\textsc V$ is equivalent to the condition
\be\label{commnabla}
[\hat{\nabla}\stackrel{\circ}{,}\hat{f}]={\hat X}[f]\in\hat{\mathcal A}(\textsc{V})\,, \qquad\forall f\in{\mathcal A}\,,
\ee
where there is a slight abuse of notation on the right-hand-side in the sense that it is a scalar endomorphism but it has been identified with the element ${\hat X}[f]$ of ${\mathcal A}$. 
An equivalent  definition of a covariant derivative is as a first-order differential operator $\hat{\nabla}\in{\mathcal D}_{\mathcal A}^1(\textsc{V})$ such that all its commutators with scalar endomorphisms $\hat{f}\in\hat{\mathcal A}(\textsc{V})$ are scalar endomorphisms $[\hat{\nabla}\stackrel{\circ}{,}\hat{f}]\in\hat{\mathcal A}(\textsc{V})$. Let us recall that, for a generic first-order differential operator $\hat D$, the commutator with a scalar endomorphism is an $\mathcal A$-linear endomorphism $[\hat{D}\stackrel{\circ}{,}\hat{f}]\in\text{End}_{\mathcal A}(\textsc{V})$. 

\vspace{3mm}
\noindent{\small\textbf{Example (Covariant derivative along a vector field)\,:} Consider a vector bundle $\mathbb{V}$ over a manifold $M$ and $\textsc{V}=\Gamma(\mathbb{V})$ its vector space of global sections. A covariant derivative on the ${\mathcal C}^\infty(M)$-module 
$\Gamma(\mathbb{V})$ along the derivation ${\hat X}\in{\mathfrak{X}}(M)$ is a linear map
\be\label{Kconn}
\nabla_{\hat{X}}\,:\,\Gamma(\mathbb{V})\to \Gamma(\mathbb{V})\,:\,\sigma\mapsto\nabla_{\hat{X}}\sigma
\ee
obeying to the following Leibniz rule:
\be\label{Leibnitzalong}
\nabla_{\hat{X}} (f\cdot\sigma)\,=\,{\hat{X}}[f]\cdot \sigma\,+\,f\cdot\nabla_{\hat{X}}\,\sigma\,,\qquad\forall f\in{\mathcal C}^\infty(M)\,,\quad\forall \sigma\in\Gamma(\mathbb{V})\,,
\ee
which can be phrased in operatorial way as
\be
\nabla_{\hat{X}}\circ\hat{f}=\hat{f}\circ\nabla_{\hat{X}}+\hat{X}[f]\,,\qquad\forall f\in{\mathcal C}^\infty(M)\,.
\ee 
It will be called a \textbf{covariant derivative acting on the vector bundle} $\mathbb{V}$ and
\textbf{along the vector field} $\hat{X}$. The word ``covariant'' indicates that it maps sections of $\mathbb{V}$ to sections of $\mathbb{V}$, the word ``derivative'' indicates that it is a derivation of the associative algebra $\otimes_{{\mathcal C}^\infty(M)}\Gamma(\mathbb{V})=\Gamma(\otimes\mathbb{V})$ of ``tensor fields'', as will be discussed below. It follows from the previous remarks that these widespread objects in theoretical physics can be interpreted geometrically as infinitesimal automorphisms of the underlying vector bundle $\mathbb{V}$, a fact which is rarely emphasised.\footnote{However, there are important exceptions, see \textit{e.g.} \cite{Kosmann-Schwarzbach} where this interpretation is mentioned.}
}

Finally, a corollary of a theorem of Bourbaki\footnote{See Proposition 14 in \cite[Subsection III.10.9]{Bourbaki} or Proposition 1.5 in \cite{Kosmann-Schwarzbach}.} is that the covariant derivatives on an $\mathcal A$-module $\textsc V$ are those endomorphisms of the vector space $\textsc V$ which can be extended to derivations of the tensor algebra $\otimes_{\mathcal A}(\textsc{V})$ that preserve the $\mathbb N$-grading of the latter algebra by the tensor degree. In this sense, the generalised Leibnitz rule \eqref{forallsigma} is literally the Leibnitz rule for the corresponding derivation $\hat\nabla\in\mathfrak{der}\big(\,\otimes_{\mathcal A}(\textsc{V})\,\big)$ with $\hat\nabla|_{\mathcal A}=\hat{X}\in\mathfrak{der}({\mathcal A})$, and the anchor is simply the restriction to the commutative subalgebra $\otimes^0_{\mathcal A}(\textsc{V})\cong{\mathcal A}$ of the tensor algebra. This extension is unique so a covariant derivatives on an $\mathcal A$-module $\textsc V$ can be defined equivalently as a derivation of the tensor algebra $\otimes_{\mathcal A}(\textsc{V})$ that preserves the tensor degree.

\vspace{3mm}
\noindent{\small\textbf{Example (Covariant derivatives as derivations)\,:} Let $\mathbb{V}$ be a vector bundle over a manifold $M$ and $\textsc{V}=\Gamma(\mathbb{V})$ its vector space of sections. The associative algebra $\Gamma(\otimes\mathbb{V})\cong\otimes_{{\mathcal C}^\infty(M)}\Gamma(\mathbb{V})$ is $\mathbb N$-graded by the tensor degree and $\otimes^r\mathbb{V}$ is the bundle of finite-rank whose sections are tensor fields of rank $r$.
A covariant derivative $\nabla_{\hat{X}}$ on the ${\mathcal C}^\infty(M)$-module $\Gamma(\mathbb{V})$ along the derivation ${\hat X}\in{\mathfrak{X}}(M)$ can be thought as a derivation of the associative algebra $\Gamma(\otimes\mathbb{V})\cong\otimes_{{\mathcal C}^\infty(M)}\Gamma(\mathbb{V})$ of tensor fields preserving their rank  (hence as a very particular type of vector fields on the vector bundle $\mathbb{V}$). This also justifies the terminology ``covariant derivative'' since they indeed are ``derivatives'' (they are derivation) that are ``covariant'' (they map tensor fields to tensor fields of the same type).
}

\vspace{3mm}
The Lie subalgebra $\mathfrak{cder}_{\mathcal A}(\textsc{V})\subset{\mathfrak{D}}_{\mathcal A}(\textsc{V})$ of all covariant derivatives $\hat{\nabla}$ on the $\mathcal A$-module $\textsc V$ is endowed with a structure of Lie-Rinehart algebra over $\mathcal A$ via the surjective anchor 
\be\label{surjanchorsigma}
\rho_{\mathfrak{cder}({\mathcal A})}\,:\,\mathfrak{cder}_{\mathcal A}({\textsc V})\twoheadrightarrow \mathfrak{der}({\mathcal A})\,:\,\hat{\nabla}\mapsto \hat{X}\,,
\ee
which maps 
a covariant derivative $\hat{\nabla}$ along a derivation ${\hat X}$ 
to the latter derivation. The Lie-Rinehart algebra $\mathfrak{cder}_{\mathcal A}(\textsc{V})$ spanned by the covariant derivatives will be called the \textbf{Atiyah algebra of the left module} $\textsc V$ \textbf{over the commutative algebra} $\mathcal A$.
 
The $\mathcal A$-Lie algebra $\mathfrak{gl}_{\mathcal A}(\textsc{V})\subset\mathfrak{cder}_{\mathcal A}(\textsc{V})$ of $\mathcal A$-module endomorphisms is a Lie-Rinehart ideal of the Lie-Rinehart algebra of covariant derivatives. Moreover, the quotient of the latter by the former is isomorphic to the Lie-Rinehart algebra of derivations, 
\be
\mathfrak{der}({\mathcal A})\cong \mathfrak{cder}_{\mathcal A}(\textsc{V})\,/\,\mathfrak{gl}_{\mathcal A}(\textsc{V})\,.
\ee
In fact, the Atiyah sequence of the Lie-Rinehart algebras of covariant derivatives, generalising \eqref{sxsD1A}, is
\be\label{shortexactvectLieR}
0\to\mathfrak{gl}_{\mathcal A}(\textsc{V})\hookrightarrow\mathfrak{cder}_{\mathcal A}(\textsc{V})\twoheadrightarrow\mathfrak{der}({\mathcal A})\to 0\,.
\ee
In other words, the Atiyah algebra of covariant derivatives is an extension of the Lie-Rinehart algebra of derivations by the $\mathcal A$-Lie algebra of $\mathcal A$-linear endomorphisms.

\vspace{3mm}
\noindent{\small\textbf{Example (Atiyah algebra of the commutative algebra)\,:} Consider the commutative algebra $\mathcal A$ as a module over itself. Covariant derivatives on $\mathcal A$ ($=\textsc{V}$) identify with first-order differential operators, hence $\mathfrak{cder}_{\mathcal A}({\mathcal A})\,={\mathfrak{D}}^1({\mathcal A})$ and $\mathfrak{gl}_{\mathcal A}({\mathcal A})\,=\,{\mathfrak{D}}^0({\mathcal A})$. Note that the Lie-Rinehart algebra ${\mathfrak{D}}^1({\mathcal A})$ of first-order operators on $\mathcal A$ is indeed an extension of the Lie-Rinehart algebra $\mathfrak{der}({\mathcal A})$ of derivations by the Abelian $\mathcal A$-Lie algebra ${\mathfrak{D}}^0({\mathcal A})$ of zeroth-order differential operators.
}

\vspace{3mm}
\noindent{\small\textbf{Example (Atiyah algebroid of a vector bundle)\,:} In the case where $\textsc{V}$ is a finite-rank locally-free ${\mathcal C}^\infty(M)$-module, \textit{i.e.} it is the section space $\textsc{V}=\Gamma(\mathbb{V})$ of a vector bundle $\mathbb{V}$ over $M$,
the corresponding Atiyah algebra is called the \textbf{Atiyah algebra of the vector bundle $\mathbb{V}$} and will be denoted $\mathfrak{CD}^1(\mathbb{V}):=\mathfrak{cder}_{{\mathcal C}^\infty(M)}\big(\,\Gamma(\mathbb{V})\,\big)$ for reasons which will become clear later.
The Atiyah algebra $\mathfrak{CD}^1(\mathbb{V})$ and its Lie-Rinehart ideal $\mathfrak{gl}_{{\mathcal C}^\infty(M)}\big(\,\Gamma(\mathbb{V})\,\big)$ respectively define a transitive Lie algebroid denoted $CD^1\mathbb{V}$ and a Lie algebra bundle denoted $\mathfrak{gl}(\mathbb{V})$ (with the general linear algebra $\mathfrak{gl}(V)$ as typical fibre). They are respectively called the \textbf{Atiyah algebroid} and the \textbf{general linear algebroid} of the vector bundle $\mathbb{V}$. The sections of the former and latter are, respectively covariant derivatives on the vector bundle and local endomorphisms of the fibre, \textit{i.e.} $\Gamma\big(CD^1\mathbb{V}\big)=\mathfrak{CD}^1(\mathbb{V})$ and $\Gamma\big(\,\mathfrak{gl}(\mathbb{V})\,\big)=\mathfrak{gl}_{{\mathcal C}^\infty(M)}\big(\,\Gamma(\mathbb{V})\,\big)$.
The Atiyah algebroid of a vector bundle over $M$ is an extension of the tangent bundle $TM$ by the general linear algebroid. More precisely, the \textbf{Atiyah sequence of the vector bundle} $\mathbb{V}$ is the short exact sequence of morphisms of Lie algebroids over $M$
\be
0\to \mathfrak{gl}(\mathbb{V})\stackrel{i_1}{\hookrightarrow} CD^1\mathbb{V}\stackrel{\rho_1}{\twoheadrightarrow}TM\to 0\,.
\label{shortexactvectbdAt}
\ee
It is the Lie algebroid version of the short exact sequence \eqref{shortexactvectLieR} of Lie-Rinehart algebras. 
}

\vspace{3mm}

This purely algebraic definition of the Atiyah algebroid of a vector bundle admits an equivalent definition as the Atiyah algebroid of the corresponding frame bundle. The latter is a principal bundle with general linear group as structure group. In this sense, the Atiyah algebroid of a vector bundle is a particular case of Atiyah algebroid of a principal bundle. The same is true for linear connections which can be seen as principal connections on the frame bundle. However, the algebraic constructions are somewhat more down-to-earth in the sense that they are based on effective tools for performing actual computations (like computing covariant derivatives, etc). 

\subsection{Linear connections on vector bundles}\label{linconvectbundl}

\subsubsection{Koszul connections on vector bundles}

A splitting of $\mathcal A$-module morphisms
\be\label{splitLRalg}
0\leftarrow\mathfrak{gl}_{\mathcal A}(\textsc{V})\twoheadleftarrow\mathfrak{cder}_{\mathcal A}(\textsc{V})\stackrel{\nabla}{\hookleftarrow}\mathfrak{der}({\mathcal A})\leftarrow 0\,.
\ee
of the Atiyah sequence \eqref{shortexactvectLieR} of an $\mathcal A$-module $\textsc V$ is called a \textbf{connection on the} $\mathcal A$-\textbf{module} $\textsc V$. It is equivalent to an $\mathcal A$-linear section of the surjective anchor \eqref{surjanchorsigma}, \textit{i.e.} an injective morphism of  $\mathcal A$-modules
\be\label{nablaLRalg}
\nabla_\bullet\,:\,\mathfrak{der}({\mathcal A})\hookrightarrow\mathfrak{cder}_{\mathcal A}(\textsc{V})\,:\,\hat{X}\mapsto\nabla_{\hat{X}}
\ee
from the Lie-Rinehart algebra $\mathfrak{der}({\mathcal A})$ of derivations on the commutative algebra $\mathcal A$
to the Atiyah algebra $\mathfrak{cder}_{\mathcal A}(\textsc{V})$ of covariant derivatives on the $\mathcal A$-module $\textsc V$. 
Firstly, the condition that \eqref{nablaLRalg} is section of the surjective anchor simply means that it maps a derivation $\hat{X}$ to a covariant derivative $\nabla_{\hat{X}}$ along the derivation $\hat{X}$.
Secondly, the condition that \eqref{nablaLRalg} is a morphism of $\mathcal A$-modules simply says that it is $\mathcal A$-linear (\textit{i.e.} $\nabla_{f\cdot\hat{X}}=\hat{f}\circ \nabla_{\hat{X}}$ for all $f\in\mathcal A$).
A connection on the $\mathcal A$-module $\textsc V$ is also equivalent to a decomposition of the Lie-Rinehart algebra $\mathfrak{cder}_{\mathcal A}(\textsc{V})$ of covariant derivatives as a direct sum of $\mathcal A$-modules:
\be\label{cderdecomp}
\mathfrak{cder}_{\mathcal A}(\textsc{V})\cong\mathfrak{gl}_{\mathcal A}(\textsc{V})\oplus_{{}_{\mathcal A}}\mathfrak{der}({\mathcal A})\,.
\ee
This decomposition is obviously not canonical, except perhaps in the degenerate case where the module is trivial, in which case there exists a canonical flat connection and the above sum is a semidirect sum of Lie-Rinehart algebras (see Subsection \ref{flatconn}).

\vspace{3mm}
\noindent{\small\textbf{Example (Linear connection)\,:} A \textbf{Koszul connection} on the vector bundle $\mathbb{V}$ over $M$ is a connection on the left ${\mathcal C}^\infty(M)$-module $\Gamma(\mathbb{V})$ of its global sections, \textit{i.e.} a left ${\mathcal C}^\infty(M)$-module morphism 
\be\label{nablabullet}
\nabla_\bullet\,:\,{\mathfrak{X}}(M)\hookrightarrow \mathfrak{CD}^1({\mathbb{V}})\,:\,\hat{X}\mapsto \nabla_{\hat{X}}
\ee
where $\nabla_{\hat{X}}$ is a covariant derivative \eqref{Kconn} on $\mathbb{V}$
along the vector field $\hat{X}$.
The ${\mathcal C}^\infty(M)$-linearity is the property
\be
\nabla_{f\cdot\hat{X}}\sigma=f\cdot \nabla_{\hat{X}}\sigma\,,\qquad\forall f\in{\mathcal C}^\infty(M)\,,\quad\forall \sigma\in\Gamma(\mathbb{V})\,,
\ee
which can be phrased in operatorial way as
\be
\nabla_{\hat{f}\circ\hat{X}}=\hat{f}\circ\nabla_{\hat{X}}\,,\qquad\forall f\in{\mathcal C}^\infty(M)\,.
\ee
From the point of view of symmetries, a Koszul connection allows to lift infinitesimal diffeomorphisms of the base manifold (\textit{i.e.} vector fields) to infinitesimal symmetries of the vector bundle (\textit{i.e.} covariant derivatives) in a way compatible with the linear fibration. This device is obviously not canonical, except perhaps in the degenerate case where the vector bundle is trivial (in which case there exists a canonical flat Koszul connection). In the simplest case of the unit bundle  $\mathbb{I}_M=M\times\mathbb R$, the canonical flat Koszul connection is the identity (it maps vector fields to themselves).
Note that the ${\mathcal C}^\infty(M)$-linearity is a crucial property of a Koszul connection. Geometrically, it means that the connection is ``directional'' in that it induces a corresponding morphism of Lie algebroids from the tangent bundle $TM$ to the Atiyah bundle of the vector bundle $\mathbb{V}$. This last property is why the Lie derivative fails to provide a (would-have-been canonical) Koszul connection on the tangent bundle: it is \textit{not} directional.
}

\vspace{3mm}
\noindent{\small\textbf{Example (Mackenzie connection)\,:} In terms of Lie algebroids, a Koszul connection on a vector bundle $\mathbb{V}$ is 
a linear splitting 
\be\label{linsplittingalgebroidAtvect}
0\leftarrow \mathfrak{gl}(\mathbb{V}){\twoheadleftarrow}CD^1\mathbb{V}\stackrel{\nabla}{\hookleftarrow}TM\leftarrow 0\,.
\ee
of the Atiyah sequence \eqref{shortexactvectLieR} of the vector bundle $\mathbb{V}$ (since the $C^\infty(M)$-linear map \eqref{nablabullet} is the counterpart of the fibrewise-linear map $\nabla:TM\hookrightarrow CD^1\mathbb{V}$). Similarly,
a seminal insight of Mackenzie is that Ehresmann's definition of a connection on a principal $H$-bundle $P$ is equivalent \cite[App.A]{Mackenzie} to a linear splitting 
\be
0\leftarrow \frac{VP}{H}\stackrel{\omega}{\twoheadleftarrow}\frac{TP}{H}\stackrel{\gamma}{\hookleftarrow}T\frac{P}{H}\leftarrow 0\,,
\label{shortexactvectAt2}
\ee
of the Atiyah sequence \eqref{Asequ} of the principal $H$-bundle $P$. The map $\gamma:T\frac{P}{H}\hookrightarrow \frac{TP}{H}$ is the horizontal lift of tangent vectors on the base manifold $P/H$ to horizontal tangent vectors on the total space $P$. The map $\omega:\frac{TP}{H}\twoheadrightarrow \frac{VP}{H}\cong P\times_{ad_H} \mathfrak{h}$ defines the $H$-equivariant $\mathfrak{h}$-valued one-form on $P$ in terms of which principal connections are traditionally formulated.
More generally, a linear splitting 
\be\label{linsplittingalgebroid}
0\leftarrow \text{Ker}\,\rho\stackrel{\omega}{\twoheadleftarrow}\mathbb{A}\stackrel{\gamma}{\hookleftarrow}TM\leftarrow 0\,.
\ee
of the Atiyah sequence \eqref{transitivLiealgebro} of a transitive Lie algebroid $\mathbb{A}$ will be called a \textbf{Mackenzie connection on the transitive Lie algebroid} $\mathbb{A}$. The splitting is linear in the sense that the arrows in \eqref{linsplittingalgebroid} are morphisms of vector bundles over $M$, but not necessarily morphisms of Lie algebroids. 
The motivation behind this generalisation is that it unifies linear and principal connections into a single framework: they are both seen as Mackenzie connections on a transitive Lie algebroid (in such case, the Atiyah bundle of a vector or principal bundle, respectively). A Mackenzie connection is flat if $\gamma$ is a morphism of Lie algebroids. Any transitive Lie algebroid over a contractible base manifold admits a flat Mackenzie connection \cite[Theorem IV.4]{Mackenzie}.
}
\vspace{3mm}

The notion of Mackenzie connections on a transitive Lie algebroid is a generalisation of Koszul connections on a vector bundle.
Another possible generalisation of the latter is to consider a morphism $D:\mathbb{A}\to CD^1\mathbb{V}$ of vector bundles over $M$
from a Lie algebroid $\mathbb{A}$ to the Atiyah algebroid $CD^1\mathbb{V}$ of the vector bundle $\mathbb{V}$ compatible with the anchors (\textit{i.e.} $\bar\pi\circ D=\rho$, where $\rho$ is the anchor of $\mathbb{A}$). This is called a \textbf{Lie algebroid connection of} $\mathbb{A}$ (or $\mathbb{A}$-connection) \textbf{on the vector bundle} $\mathbb{V}$.\footnote{They were first introduced in \cite{Fernandes:2002} via a definition in terms of horizontal lifts (shown there to be equivalent to the present one).}

\vspace{3mm}
\noindent{\small\textbf{Example (Koszul connection)\,:} In the particular case where the Lie algebroid is the tangent bundle (\textit{i.e.} $\mathbb{A}=TM$), the notion of $TM$-connection $\nabla:TM\to CD^1\mathbb{V}$ is nothing but a Koszul connection on $\mathbb{V}$. More generally, the composition $\nabla_{\bullet}:=D_\bullet\circ\gamma$ of a Mackenzie connection $\gamma:TM\hookrightarrow\mathbb{A}$ on a transitive Lie algebroid $\mathbb{A}$ with an $\mathbb{A}$-connection $D_\bullet:\mathbb{A}\to CD^1\mathbb{V}$ on $\mathbb{V}$ is a Koszul connection on $\mathbb{V}$.
}

\vspace{3mm}
\noindent{\small\textbf{Example (Partial connection)\,:} If the Lie algebroid is an involutive distribution on $M$ (\textit{i.e.} $\mathbb{A}\subset TM$), the notion of $\mathbb{A}$-connection $D:\mathbb{A}\to CD^1\mathbb{V}$ is sometimes called a \textbf{partial connection} on $\mathbb{V}$ (see \textit{e.g.} \cite{Kamber:1975}).\footnote{The terminology stresses that the covariant derivatives are only defined along vectors which are tangent to the leaves of the foliation} Equivalently, the composition $D_{\bullet}:=\nabla_\bullet\circ i$ of an injective anchor $i:\mathbb{A}\hookrightarrow TM$ with a Koszul connection $\nabla_\bullet:TM\hookrightarrow CD^1\mathbb{V}$ is a partial connection.
}

\vspace{3mm}
\noindent{\small\textbf{Remark:} Even more generally, one may consider a Lie-Rinehart algebra extension $\mathfrak{G}$ {of} $\mathfrak{H}$ {by} $\mathfrak{I}$ defined by the short exact sequence \eqref{shortexactLRalg} of morphisms of Lie-Rinehart algebras over $\mathcal A$\,. A splitting \eqref{splittingLRalg} where all arrows are $\mathcal A$-module morphisms is sometimes called a $\mathfrak{G}$-\textbf{connection} \cite{huebschmann1997extensions}. A $\mathfrak{G}$-connection is guaranteed to exist if $\mathfrak{H}$ is a projective $\mathcal A$-module (since the short exact sequence always splits). In particular, any short exact sequence of Lie algebroids always splits (since Lie algebroids correspond to the case of Lie-Rinehart algebras which are projective ${\mathcal C}^\infty(M)$-modules of finite rank) and each such splitting defines a  generalised connection.\footnote{For instance, a Mackenzie connection always exist for any transitive Lie algebroid.} It is in this broad sense that the theory of Lie algebroid extensions coincides with the theory of (generalised) connections on principal and vector bundles.
}

\subsubsection{Flat connections as semidirect sum decompositions}
\label{flatconn}

A connection on an $\mathcal A$-module $\textsc V$ is flat iff the section \eqref{nablaLRalg} is a morphism of Lie algebras, \textit{i.e.} 
\be
[\,\nabla_{\hat{X}}\,,\,\nabla_{\hat{Y}}\,]=\nabla_{[\hat{X},\hat{Y}]}
\ee
for all $\hat{X},\hat{Y}\in\mathfrak{der}({\mathcal A})$.
Equivalently, it is flat iff it makes 
the Atiyah algebra $\mathfrak{cder}_{\mathcal A}(\textsc{V})$ of covariant derivatives a Lie-Rinehart split extension of the Lie-Rinehart algebra $\mathfrak{der}({\mathcal A})$ of derivations by the $\mathcal A$-Lie algebra $\mathfrak{gl}_{\mathcal A}(\textsc{V})$ of $\mathcal A$-linear endomorphisms.
For a flat connection, the image of the section \eqref{nablaLRalg},
\be
\nabla_{\mathfrak{der}({\mathcal A})}\,=\,\{\,\nabla_{\hat{X}}\in\mathfrak{cder}_{\mathcal A}(\textsc{V})\,|\,\hat{X}\in\mathfrak{der}({\mathcal A})\,\}\,,
\ee
 is a Lie-Rinehart subalgebra of the Atiyah algebra $\mathfrak{cder}_{\mathcal A}(\textsc{V})$ spanned by covariant derivatives. This Lie-Rinehart subalgebra $\nabla_{\mathfrak{der}({\mathcal A})}\subset\mathfrak{cder}_{\mathcal A}(\textsc{V})$ is, by construction, isomorphic to the Lie-Rinehart algebra $\mathfrak{der}({\mathcal A})$ of derivations.
In this sense, the previous decomposition \eqref{cderdecomp} (as direct sum of $\mathcal A$-modules) becomes a stronger decomposition (as semidirect sum of Lie-Rinehart algebras)\,:
\be\label{cderniplus}
\mathfrak{cder}_{\mathcal A}(\textsc{V})\cong\mathfrak{der}({\mathcal A})\inplus_{{}_\nabla}\mathfrak{gl}_{\mathcal A}(\textsc{V})\,.
\ee

\vspace{3mm}
\noindent{\small\textbf{Example (Flat linear connection)\,:} A Koszul connection \eqref{nablabullet} on a vector bundle $\mathbb{V}$ over $M$ is flat iff it is a morphism of Lie algebras from the Lie-Rinehart algebra of vector fields on $M$ into the the Lie-Rinehart algebra of covariant derivatives on $\mathbb{V}$. It is equivalent to a decomposition of the Atiyah algebra $\mathfrak{CD}^1({\mathbb V})$ of the vector bundle as the semidirect sum of the Lie-Rinehart algebra of vector fields on the base by the ${\mathcal C}^\infty(M)$-Lie algebra of local endomorphisms, 
\be
\mathfrak{CD}^1({\mathbb V})\cong \mathfrak{X}(M)\inplus_{{}_\nabla}\Gamma\big(\,\mathfrak{gl}({\mathbb V})\,\big)\,.
\ee
From the Lie algebroid point of view, a flat Koszul connection makes the Atiyah bundle $CD^1\mathbb{V}$ into a split extension of the tangent bundle $TM$ by the general linear algebroid $\mathfrak{gl}(\mathbb{V})$.
}

\vspace{3mm}
\noindent{\small\textbf{Example (Lie-Rinehart algebra of first-order differential operators)\,:} In the particular case where the module identifies with the commutative algebra, there is a canonical flat connection \eqref{sxsD1Asplitting} on the $\mathcal A$-module ${\mathcal A}$. This is in agreement with the canonical decomposition 
\be
{\mathfrak{D}}^1({\mathcal A})\cong\mathfrak{der}({\mathcal A})\inplus{\mathfrak{D}}^0({\mathcal A})
\ee
of the Lie-Rinehart algebra of first-order differential operators on the commutative algebra $\mathcal A$ as a semidirect sum of the Abelian ideal ${\mathfrak{D}}^0({\mathcal A})\subset{\mathfrak{D}}^1({\mathcal A})$ of zeroth-order differential operators and the Lie-Rinehart subalgebra $\mathfrak{der}({\mathcal A})\subset{\mathfrak{D}}^1({\mathcal A})$ of its derivations, since
$\mathfrak{cder}_{\mathcal A}({\mathcal A})\,={\mathfrak{D}}^1({\mathcal A})$
and $\mathfrak{gl}_{\mathcal A}({\mathcal A})\,=\,{\mathfrak{D}}^0({\mathcal A})\cong\mathcal A$.
}

\subsection{Flat linear connections as representations of Lie algebroids}

Representations of Lie algebroids are an abstract generalisation of flat Koszul connections.
 
\subsubsection{Representations of Lie-Rinehart algebras}

Let $\mathfrak{L}$ be a Lie-Rinehart algebra over $\mathcal A$ with anchor $\rho$ and let $\textsc{V}$ be an $\mathcal A$-module.

A \textbf{representation of a Lie-Rinehart algebra} $\mathfrak{L}$ \textbf{on a left module} $\textsc V$ over the same commutative algebra $\mathcal A$ is a morphism
\be\label{nablamorph}
\nabla\,:\,\mathfrak{L}\to\mathfrak{cder}_{\mathcal A}(\textsc{V})\,:\,\hat{D}\mapsto\hat{\nabla}\,.
\ee
of Lie-Rinehart algebras over $\mathcal A$ from $\mathfrak{L}$ to $\mathfrak{cder}_{\mathcal A}(\textsc{V})$. The compatibility of the respective anchors \eqref{surjanchorsigma} and $\rho$ reads $\rho_{\mathfrak{cder}({\mathcal A})}\circ\nabla=\rho$, which simply means that an element $\hat{D}\in\mathfrak{L}$ is mapped to a covariant derivative $\hat{\nabla}$ along the derivation $\hat{X}=\rho(\hat{D})\in\mathfrak{der}({\mathcal A})$.

A module $\textsc V$ of the commutative algebra $\mathcal A$ carrying a representation of a Lie-Rinehart algebra $\mathfrak{L}$ over $\mathcal A$ is called a \textbf{module of the Lie-Rinehart algebra} $\mathfrak{L}$. It means that it is a module of (both) the commutative algebra $\mathcal A$ and the Lie algebra $\mathfrak{L}$, such that their respective actions on $\textsc V$ are compatible.

\vspace{3mm}
\noindent{\small\textbf{Example (Free module)\,:} There is a canonical representation of the Lie-Rinehart algebra $\mathfrak{der}({\mathcal A})$ of derivations of $\mathcal A$ on any free ${\mathcal A}$-module generated by a vector space $V$, called the \textbf{trivial representation}, where the derivations simply act on the first factor (the commutative algebra $\mathcal A$) of the free ${\mathcal A}$-module $\textsc{V}={\mathcal A}\otimes V$ and do not touch the second factor (the vector space $V$), \textit{i.e.} \be
\nabla_{\hat{X}}\,(f\otimes v):=\hat{X}[f]\otimes v\,,\qquad \forall \hat{X}\in\mathfrak{der}({\mathcal A}),\quad\forall f\in\mathcal{A}, \quad\forall v\in V\,.
\ee
}

\noindent{\small\textbf{Example (Bott representation)\,:} Consider a Lie pair $\mathfrak{H}\subset\mathfrak{G}$, \textit{i.e.} a Lie-Rinehart subalgebra $\mathfrak{H}$ of $\mathfrak{G}$ over the same commutative algebra $\mathcal A$.
The quotient $\mathfrak{G}/\mathfrak{H}$ is a module of the Lie-Rinehart algebra $\mathfrak{H}$. 
In fact, canonically there is a surjective morphism 
$\pi:\mathfrak{G}\twoheadrightarrow\mathfrak{G}/\mathfrak{H}:\hat{G}\mapsto [\hat{G}]$ of $\mathcal A$-modules as well as a morphism of Lie-Rinehart algebras over $\mathcal A$,
\be\label{nablamorphBott}
\nabla\,:\,\mathfrak{H}\to\mathfrak{cder}_{\mathcal A}(\mathfrak{G}/\mathfrak{H})\,:\,\hat{H}\mapsto\nabla_{\hat{H}}\,,
\ee
defined by
\be
\nabla_{\hat{H}}\pi(\hat{G})\,:=\,\pi\big(\,[\hat{H},\hat{G}]\,\big)\,.
\ee
The latter morphism is sometimes\footnote{\textit{cf.} Footnote \ref{prevfootn}.} called the \textbf{Bott representation} of the Lie-Rinehart $\mathfrak{H}$ over the $\mathcal A$-module $\mathfrak{G}/\mathfrak{H}$.
The Bott representation vanishes identically ($\nabla=0$) iff $\mathfrak{H}$ is a Lie-Rinehart ideal of $\mathfrak{G}$.
}

\vspace{3mm}

The Atiyah functor $CD^1$ mapping a vector bundle $\mathbb{V}$ to its Atiyah Lie algebroid $CD^1\mathbb{V}$
is the algebroid generalisation of the functor $\mathfrak{gl}$ mapping a vector space $V$ to its general linear Lie algebra $\mathfrak{gl}(V)$.
By analogy with the definition of representation of a Lie algebra $\mathfrak{g}$ on the vector space $V$ as a morphism $\mathfrak{g}\to\mathfrak{gl}(V)$ of Lie algebras, it is natural to define a \textbf{representation of a Lie algebroid} $\mathbb{A}$ \textbf{on the vector bundle} $\mathbb{V}$ (over the same base $M$) as a morphism $\mathbb{A}\to CD^1\mathbb{V}$ of Lie algebroids over $M$.
It is the geometrical analogue of a representation of a Lie-Rinehart algebra $\mathfrak{L}$ on a module $\textsc V$ (over the same commutative algebra $\mathcal A$) since it leads to a representation of the Lie-Rinehart algebra $\Gamma(\mathbb{A})$ ($=\mathfrak{L}$) over the structure algebra $C^\infty(M)$ ($=\mathcal A$) on the left $C^\infty(M)$-module $\Gamma(\mathbb{V})$ ($=\textsc V$), \textit{i.e.} a Lie-Rinehart algebra morphism $\nabla\,:\,\Gamma(\mathbb{A})\to\mathfrak{CD}^1(\mathbb{V})$.

\vspace{3mm}
\noindent{\small\textbf{Example (Trivial vector bundle)\,:} A free $C^\infty(M)$-module $C^\infty(M)\otimes V$ generated by a vector space $V$ is the space $\Gamma(\mathbb{V})$ of sections of a trivial vector bundle $\mathbb{V}=M\times V$ over $M$ with fibre $V$. Its sections are $V$-valued functions on $M$ and vector fields on $M$ act on them in a natural way, via the trivial representation of the Lie-Rinehart algebra $\mathfrak{X}(M)=\Gamma(TM)$ of vector fields on the section space $\Gamma(M\times V)=C^\infty(M)\otimes V$ where $\mathfrak{X}(M)$ acts only on the first factor $C^\infty(M)$. Accordingly, this defines the trivial representation of the tangent bundle $TM$ on any trivial vector bundle over $M$.
}

\subsubsection{Flat connections as representations}

The definition of representation of Lie-Rinehart algebras allows to interpret a flat connection on an $\mathcal A$-module $\textsc V$ as a faithful representation $\nabla:\mathfrak{der}({\mathcal A})\hookrightarrow\mathfrak{cder}_{\mathcal A}(\textsc{V})$ of the Lie-Rinehart algebra $\mathfrak{der}({\mathcal A})$ of derivations on the $\mathcal A$-module $\textsc V$. Indeed, the compatiblity of their anchors reads $\rho_{\mathfrak{cder}({\mathcal A})}\circ\nabla=id_{\mathfrak{der}({\mathcal A})}$ and implies that $\nabla$ is a section of the anchor \eqref{surjanchorsigma} of the Atiyah algebra.

\vspace{3mm}
\noindent{\small\textbf{Example (Flat linear connection)\,:} A flat Koszul connection on a vector bundle $\mathbb{V}$ over $M$ defines a faithful representation, of the Lie-Rinehart algebra ${\mathfrak{X}}(M)$ of base vector fields, on the $C^\infty(M)$-module $\Gamma(\mathbb{V})$ of sections of the vector bundle. From the Lie algebroid point of view, a flat Koszul connection is a faithful representation of the tangent bundle $TM$ (seen as a Lie algebroid) on the vector bundle $\mathbb{V}$.
}

\vspace{5mm}
\begin{framed}
\begin{center}
\textbf{Algebraic formulation of flat connections}
\end{center}

\noindent
Given a left $\mathcal A$-module $\textsc V$, the following notions are equivalent:
\begin{enumerate}
 \item a flat connection on the $\mathcal A$-module $\textsc V$,
 \item a faithful representation $\nabla:\mathfrak{der}({\mathcal A})\hookrightarrow\mathfrak{cder}_{\mathcal A}(\textsc{V})$ of the Lie-Rinehart algebra $\mathfrak{der}({\mathcal A})$ of derivations, on the $\mathcal A$-module $\textsc V$,
 \item a Lie-Rinehart split extension of the Lie-Rinehart algebra of derivations $\mathfrak{der}({\mathcal A})$ by the $\mathcal A$-Lie algebra $\mathfrak{gl}_{\mathcal A}(\textsc{V})$ of $\mathcal A$-linear endomorphisms, \textit{i.e.} a splitting of the Atiyah sequence such that all arrows are $\mathcal A$-module morphisms and the section $\nabla$ is a morphism of Lie-Rinehart algebras over $\mathcal A$,
 \item a decomposition of the Atiyah algebra $\mathfrak{cder}_{\mathcal A}(\textsc{V})$ of covariant derivatives on the $\mathcal A$-module $\textsc V$ as the semidirect sum of the Lie-Rinehart ideal $\mathfrak{gl}_{\mathcal A}(\textsc{V})$ of $\mathcal A$-linear endomorphisms with the Lie-Rinehart algebra $\mathfrak{der}({\mathcal A})$ of derivations.
\end{enumerate}
\vspace{3mm}
\end{framed}

\subsubsection{Adjoint representation of a Lie-Rinehart algebra}

If the $\mathcal A$-module $\textsc V$ has more structure (than being a left $\mathcal A$-module, \textit{e.g.} if it is an $\mathcal A$-Lie algebra) then it is natural to require the representation to preserve this extra structure. In particular, one will focus here on representations of Lie-Rinehart algebras on $\mathcal A$-Lie algebras and observe that they are in one-to-one correspondence with semidirect sums of Lie-Rinehart algebras.

Consider an $\mathcal A$-Lie algebra $\mathfrak{I}$. A covariant derivative $\hat{\nabla}$ on $\mathfrak{I}$ (seen as a left $\mathcal A$-module) which is also a derivation of $\mathfrak{I}$ (seen as a Lie algebra), \textit{i.e.} which is such that
\be
\hat{\nabla}[\hat{Y}_1,\hat{Y}_2]=[\hat{\nabla}\hat{Y}_1,\hat{Y}_2]+[\hat{Y}_1,\hat{\nabla}\hat{Y}_2]\,,
\ee
for all $\hat{Y}_1,\hat{Y}_2\in\mathfrak{I}$, will be called a \textbf{Lie-covariant derivative on the} $\mathcal A$\textbf{-Lie algebra} $\mathfrak{I}$.
These Lie-covariant derivates form a Lie-Rinehart subalgebra which is the intersection
\be
\overline{\mathfrak{cder}}_{\mathcal A}(\mathfrak{I})\,:=\,{\mathfrak{cder}}_{\mathcal A}(\mathfrak{I})\cap{\mathfrak{der}}(\mathfrak{I})
\ee
of the Atiyah algebra ${\mathfrak{cder}}_{\mathcal A}(\mathfrak{I})$ of covariant derivatives of $\mathfrak{I}$ and of the Lie-Rinehart algebra ${\mathfrak{der}}(\mathfrak{I})$ of derivations of $\mathfrak{I}$. 
The Lie-Rinehart algebra $\overline{\mathfrak{cder}}_{\mathcal A}(\mathfrak{I})$ of Lie-covariant derivatives will be called the \textbf{Atiyah algebra of the} $\mathcal A$\textbf{-Lie algebra} $\mathfrak{I}$.
The $\mathcal A$-linear endomorphisms of $\mathfrak{I}$ (seen as a left $\mathcal A$-module) which are also derivations of $\mathfrak{I}$ (seen as a Lie algebra) span an $\mathcal A$-Lie algebra denoted by 
\be
\overline{\mathfrak{der}}_{\mathcal A}(\mathfrak{I})\,:=\,{\mathfrak{gl}}_{\mathcal A}(\mathfrak{I})\cap{\mathfrak{der}}(\mathfrak{I})\,.
\ee
It will be called the \textbf{derivation algebra of the} $\mathcal A$\textbf{-Lie algebra} $\mathfrak{I}$.

A \textbf{representation of a Lie-Rinehart algebra} $\mathfrak{L}$ \textbf{over} $\mathcal A$ \textbf{on an} $\mathcal A$\textbf{-Lie algebra} $\mathfrak{I}$, is a morphism $\nabla\,:\,\mathfrak{L}\to\overline{\mathfrak{cder}}_{\mathcal A}(\mathfrak{I})$ of Lie-Rinehart algebras from the Lie-Rinehart algebra $\mathfrak{L}$ to the Atiyah algebra of Lie-covariant derivatives on the $\mathcal A$-Lie algebra $\mathfrak{I}$.

\vspace{3mm}
\noindent{\small\textbf{Example (Adjoint representation)\,:} Consider a Lie-Rinehart algebra $\mathfrak{L}$ over $\mathcal A$. The kernel $\text{Ker}\,\rho\subset\mathfrak{L}$ of its anchor $\rho$ is a Lie-Rinehart ideal. There is a canonical representation
\be
ad\,:\,\mathfrak{L}\to\overline{\mathfrak{cder}}_{\mathcal A}(\text{Ker}\,\rho)\,:\,\hat{X}\mapsto ad_{\hat X}
\ee
of any Lie-Rinehart algebra on the kernel of its anchor, called the \textbf{adjoint representation}. The Lie-covariant derivative $ad_{\hat X}$ is defined by
\be
ad_{\hat{X}}(\hat{Y})\,:=\,[\,\hat{X}\,,\,\hat{Y}\,]\,,\qquad\forall\hat{Y}\in\text{Ker}\,\rho\,.
\ee
It will be called an \textbf{inner covariant derivative}. 
}

\vspace{3mm}
\noindent{\small\textbf{Example (Transitive Lie algebroid):}  In terms of Lie algebroids, the previous construction implies that there exists a canonical representation of any transitive Lie algebroid $\mathbb{A}$ on its adjoint algebroid $\text{Ker}\,\rho\subset\mathbb{A}$. In the particular case of the Atiyah algebroid of a vector bundle, the covariant derivatives on the vector bundle actually acts on the associative algebra $\text{End}_{C^\infty(M)}\big(\,\Gamma(\mathbb{V})\,\big)$ of $C^\infty(M)$-linear maps. Note that it acts via maps which are covariant derivatives with respect to the $C^\infty(M)$-module structure and which are derivations with respect to the composition product.
}

\vspace{3mm}\noindent\textbf{Remark:} There is a canonical representation of any Lie algebra $\mathfrak{g}$ on itself: the adjoint representation. This remains true for $\mathcal A$-Lie algebras. However, it is \text{not} true for Lie-Rinehart algebras $\mathfrak{L}$ over $\mathcal A$ with non-trivial anchor $\rho\neq 0$. Although the adjoint map $ad:\mathfrak{L}\to\mathfrak{cder}_{\mathcal A}(\mathfrak{L})$ is a morphism of Lie algebras, it fails to be an $\mathcal A$-module morphism (since it will not be $\mathcal A$-linear when the anchor is non-trivial). 
In particular, the Lie derivative does not provide a flat Koszul connection on the tangent bundle because it fails to be $C^\infty(M)$-linear.
Nevertheless, by definition the anchor provides a canonical representation of any Lie-Rinehart algebra $\mathfrak{L}$ on its underlying commutative algebra $\mathcal A$. Moreover, there is a canonical representation of any Lie algebra $\mathfrak{g}$ on its universal enveloping algebra ${\mathcal U}(\mathfrak{g})$ via left multiplication, a property which extends to Lie-Rinehart algebras.

\vspace{3mm}
\noindent{\small\textbf{Example (Atiyah algebroid of a Lie algebra bundle):} Consider a weak Lie algebra bundle $\mathbb{L}$ over $M$, \textit{i.e.} a Lie algebroid over $M$ whose section space $\mathfrak{L}:=\Gamma(\mathbb{L})$ is a $C^\infty(M)$-Lie algebra. 
Let us denote by 
\be\label{AtiyahLAB}
\overline{\mathfrak{CD}}^1({\mathfrak{L}})\,:=\,\mathfrak{CD}^1({\mathfrak{L}})\cap{\mathfrak{der}}(\mathfrak{L})
\ee
the Lie-Rinehart subalgebra spanned by the Lie-covariant derivatives on the $C^\infty(M)$-Lie algebra bundle $\mathbb{L}$ over $M$.
One may also consider the Lie-Rinehart subalgebra
\be
\overline{\mathfrak{gl}}(\mathfrak{L})\,:=\,\mathfrak{gl}_{C^\infty(M)}({\mathfrak{L}})\cap{\mathfrak{der}}(\mathfrak{L})
\ee
spanned by the $C^\infty(M)$-linear endomorphisms which are derivations of the $C^\infty(M)$-Lie algebra $\mathfrak{L}$. 
The Lie algebroids corresponding to the Lie-Rinehart algebras $\overline{\mathfrak{CD}}^1({\mathfrak{L}})$ and $\overline{\mathfrak{gl}}(\mathfrak{L})$
will be denoted, respectively, $\overline{CD}^1\mathbb{L}$ and $\overline{\mathfrak{gl}}(\mathbb{L})$. 
The corresponding Atiyah sequence of the Lie algebra bundle $\mathbb{L}$ is the short exact sequence
\be
0\to \overline{\mathfrak{gl}}(\mathbb{L})\stackrel{i_1}{\hookrightarrow} \overline{CD}^1\mathbb{L}\stackrel{\rho_1}{\twoheadrightarrow}TM\to 0
\ee
of Lie algebroids over $M$.
In the particular case of a strong Lie algebra bundle $\mathbb{L}$ over $M$ with the Lie algebra $\mathfrak{g}$ as typical fibre, the typical fibre of the Lie algebra bundle $\overline{\mathfrak{gl}}(\mathbb{L})$ is the Lie algebra $\mathfrak{der}(\mathfrak{g})$ of derivations of the fibre.
}

\vspace{3mm}
\noindent{\small\textbf{Example (Lie connection):} Consider a $C^\infty(M)$-Lie algebra bundle $\mathbb{L}$ over $M$ and the corresponding Lie-Rinehart subalgebra \eqref{AtiyahLAB} of the Atiyah algebra for its section space $\mathfrak{L}=\Gamma(\mathbb{L})$. 
A \textbf{Lie connection} is a Koszul connection $\nabla_\bullet:{\mathfrak{X}}(M)\hookrightarrow \overline{\mathfrak{CD}}^1({\mathfrak{L}})$ on the $C^\infty(M)$-Lie algebra bundle $\mathbb{L}$ whose covariant derivatives are Lie-covariant derivatives of the $C^\infty(M)$-Lie algebra $\mathfrak L$. This means that the structure constants are covariantly constant with respect to a Lie connection.
Not surprisingly, a $C^\infty(M)$-Lie algebra bundle $\mathbb{L}$ over $M$ is a Lie algebra bundle iff it admits a Lie connection \cite[Theorem III.7.12]{Mackenzie}).
}

\subsubsection{Semidirect sums of Lie-Rinehart algebras as representations}\label{semidirectsumsLR}

If the injective arrow in the splitting \eqref{splittingLRalg} is a Lie algebra morphism, then it is called a \textbf{flat} $\mathfrak{G}$-\textbf{connection}. It is equivalent to a Lie-Rinehart algebra splitting of the short exact sequence \eqref{shortexactLRalg} of Lie-Rinehart algebra or, in other words a decomposition as a semidirect sum $\mathfrak{G}\cong\mathfrak{H}\inplus_{{}_\nabla}\mathfrak{I}$. The corresponding adjoint action of the Lie-Rinehart subalgebra $\mathfrak{H}\subset \mathfrak{G}$ on the Lie-Rinehart ideal $\mathfrak{I}\subset \mathfrak{G}$ induces a representation $\nabla$ of $\mathfrak{H}$ on $\mathfrak{I}$.

Conversely, given with a representation $\nabla$ of a Lie-Rinehart algebra $\mathfrak{H}$ on an $\mathcal A$-Lie algebra $\mathfrak{I}$, one can form the semidirect sum of the latter Lie-Rinehart algebras. To be precise, the semidirect sum $\mathfrak{G}=\mathfrak{H}\inplus_{{}_\nabla}\mathfrak{I}$ is defined as the Lie-Rinehart algebra whose:
\begin{itemize}
	\item left $\mathcal A$-module structure
is the direct sum of the $\mathcal A$-modules $\mathfrak{H}$ and $\mathfrak{I}$, \textit{i.e.} 
\be
f\cdot(\hat{X}\oplus\hat{Y})=(f\cdot\hat{X})\oplus(f\cdot\hat{Y})\,,\qquad\forall f\in{\mathcal A}\,,\,\forall\hat{X}\in\mathfrak{H}\,,\,\forall\hat{Y}\in\mathfrak{I}\,.
\ee
	\item Lie algebra structure is the semidirect sum of their Lie algebras, \textit{i.e.}
\be
[\,\hat{X}_1\oplus \hat{Y}_1\,,\,\hat{X}_2\oplus \hat{Y}_2\,]_\mathfrak{G}=[\,\hat{X}_1\,,\,\hat{X}_2\,]_\mathfrak{H}\,\oplus\,\Big(\,\nabla_{\hat{X}_1}\hat{Y}_2-\nabla_{\hat{X}_2}\hat{Y}_1+[\,\hat{Y}_1\,,\,\hat{Y}_2\,]_\mathfrak{I}\,\Big)
\ee
for all $\hat{X}_1,\hat{X}_2\in\mathfrak{H}$ and $\hat{Y}_1,\hat{Y}_2\in\mathfrak{I}$.
	\item anchor reduces to the one of $\mathfrak{H}$ ($\rho_{\mathfrak{G}}=\rho_{\mathfrak{H}}\oplus 0$), \textit{i.e.}
\be
\rho_{\mathfrak{G}}\,:\,\hat{X}\oplus\hat{Y}\mapsto\rho_{\mathfrak{H}}(\hat{X})\qquad\forall\hat{X}\in\mathfrak{H}\,,\,\forall\hat{Y}\in\mathfrak{I}\,.
\ee	
\end{itemize}

\vspace{3mm}
\noindent{\small\textbf{Example (Trivial Lie-Rinehart algebra):} Consider a free ${\mathcal A}$-module $\textsc{V}={\mathcal A}\,\otimes V$ generated by a vector space $V$. The trivial representation is the canonical representation of the Lie-Rinehart algebra $\mathfrak{der}({\mathcal A})$ of derivations on the free ${\mathcal A}$-module $\textsc{V}={\mathcal A}\,\otimes V$ that acts only on the first factor.  Moreover, the associative algebra of $\mathcal A$-linear endomorphisms is the tensor product $\text{End}_{\mathcal A}({\mathcal A}\otimes V)\cong{\mathcal A}\otimes\text{End}(V)$ of the commutative algebra $\mathcal A$ by the associative algebra $\text{End}(V)$ of endomorphisms of the underlying vector space $V$. Therefore, the $\mathcal A$-Lie algebra of $\mathcal A$-linear endomorphisms is a free ${\mathcal A}$-module $\mathfrak{gl}_{\mathcal A}({\mathcal A}\otimes V)\cong{\mathcal A}\otimes\mathfrak{gl}(V)$ generated by the general linear algebra of the underlying vector space $V$. The Lie-Rinehart algebra of covariant derivatives decomposes as the semidirect sum of the Lie-Rinehart algebra of derivations and the Lie-Rinehart ideal of $\mathcal A$-linear endomorphisms
\be\label{cdertrivialLiealg}
\mathfrak{cder}_{\mathcal A}({\mathcal A}\otimes V)\,\cong\,\mathfrak{der}({\mathcal A})\,\inplus\,\big(\,{\mathcal A}\otimes\mathfrak{gl}(V)\,\big)\,.
\ee 
More generally, the similar semidirect sum $\mathfrak{der}({\mathcal A})\inplus({\mathcal A}\otimes\mathfrak{g})$ of the Lie-Rinehart algebra $\mathfrak{der}({\mathcal A})$ of derivations with the ${\mathcal A}$-Lie algebra ${\mathcal A}\otimes\mathfrak g$ generated by the Lie algebra $\mathfrak g$ will be called the \textbf{trivial Lie-Rinehart algebra over} $\mathcal A$ \textbf{with isotropy algebra} $\mathfrak{g}$.
}

\vspace{3mm}
\noindent{\small\textbf{Example (Trivial Lie algebroid):} A simple class of examples of transitive Lie algebroids over $M$ with isotropy algebra $\mathfrak{g}$ is the semidirect sum $TM\inplus(M\times\mathfrak{g})$ of the tangent bundle $TM$ and of the trivial Lie algebra bundle over $M$ with fibre $\mathfrak{g}$.
This transitive Lie algebroid is called the \textbf{trivial Lie algebroid over} $M$ \textbf{with isotropy algebra} $\mathfrak{g}$.
As a vector bundle over $M$, it decomposes as the direct sum $TM\oplus(M\times\mathfrak{g})$. 
Its Atiyah algebra is the trivial Lie-Rinehart algebra over the structure algebra $C^\infty(M)$ with isotropy algebra $\mathfrak{g}$,
\be
\Gamma\big(\,TM\inplus(M\times\mathfrak{g})\,\big)\,=\,\mathfrak{X}(M)\,\inplus\,\big(\,C^\infty(M)\otimes\mathfrak{g}\,\big)\,.
\ee
The Atiyah sequence of a trivial Lie algebroid reads
\be\label{trivLiealgebro}
0\to M\times\mathfrak{g}\hookrightarrow TM\inplus(M\times\mathfrak{g})\twoheadrightarrow TM\to 0\,.
\ee
In other words, the adjoint algebroid of the trivial Lie algebroid over $M$ {with isotropy algebra} $\mathfrak{g}$ is the trivial Lie algebra bundle $M\times\mathfrak{g}$. A transitive Lie algebroid $\mathbb{A}$ over $M$ with isotropy algebra $\mathfrak{g}$ has the structure of a trivial Lie algebroid, \textit{i.e.} $\mathbb{A}\cong TM\inplus(M\times\mathfrak{g})$, iff it admits a flat Mackenzie connection (\textit{cf.} \cite[Theorem IV.4.1]{Mackenzie} and comments thereafter).
}

\vspace{3mm}
\noindent{\small\textbf{Example (Flat Koszul connection):} The Atiyah bundle $CD^1\mathbb{V}$ of a vector bundle $\mathbb{V}$ over $M$ with fibre $V$ has the structure of a trivial Lie algebroid $TM\inplus\big(M\times\mathfrak{gl}(V)\big)$ iff the vector bundle admits a flat Koszul connection (\textit{cf.} the comment in \cite[Sect.1]{Kosmann-Schwarzbach}).
}

\vspace{3mm}
\noindent{\small\textbf{Example (Trivial vector bundle):} Consider a trivial vector bundle $\mathbb{V}=M\times V$ over $M$ with fibre $V$. Its general linear algebroid is the trivial Lie algebra bundle $\mathfrak{gl}(M\times V)=M\times \mathfrak{gl}(V)$ over $M$ with fibre the general linear algebra $\mathfrak{gl}(V)$.
The Atiyah algebroid of the trivial vector bundle $\mathbb{V}=M\times V$ is the trivial Lie algebroid over $M$ with isotropy algebra $\mathfrak{gl}(V)$, \textit{i.e.} the Lie algebroid $TM\inplus\big(M\times\mathfrak{gl}(V)\big)$. 
These features are distinct facets of the property that trivial vector bundles are endowed with a canonical flat Koszul connection whose flat sections are simply the $V$-valued constant functions on $M$.
}

\pagebreak

\section{The universal enveloping algebra of a Lie-Rinehart algebra}\label{Assalgebras}

The definition of the universal enveloping algebra of a Lie-Rinehart algebra has to be refined with respect to the standard case of a Lie algebra. In fact, the structure of Lie-Rinehart algebra $\mathfrak{L}$ over a commutative algebra $\mathcal A$ is richer than the one of a Lie algebra $\mathfrak{g}$ over a field $\mathbb K$. In practice, the construction should take into account the $\mathcal A$-module structure of $\mathfrak{L}$.

\subsection{Universal enveloping algebras of Lie-Rinehart algebras}

\subsubsection{Universal enveloping algebras of Lie algebras}

The \textbf{universal enveloping algebra} ${\mathcal U}(\mathfrak{g})$ \textbf{of a Lie algebra} $\mathfrak{g}$ is the quotient of the tensor algebra $\otimes(\mathfrak{g})$ by the two-sided ideal generated by the identification of the commutator of generators with their Lie bracket  (\textit{i.e.} the associative ideal spanned by elements proportional to $\hat{X}\otimes\hat{Y}-\hat{Y}\otimes\hat{X}-[\hat{X},\hat{Y}]$ for some $\hat{X},\hat{Y}\in\mathfrak{g}$). 

The Poincar\'e-Birkhoff-Witt theorem asserts that the universal enveloping algebra ${\mathcal U}(\mathfrak{g})$ of the Lie algebra $\mathfrak{g}$  is an almost-commutative algebra whose associated graded algebra is isomorphic to the symmetric algebra of $\mathfrak{g}$, 
\be
\text{gr}\,{\mathcal U}(\mathfrak{g})\,\cong\,\odot(\mathfrak{g})\,.
\ee
Note that the universal enveloping algebra ${\mathcal U}(\mathfrak{g})$ of a Lie algebra $\mathfrak{g}$ is never simple, since the subalgebra ${\mathcal U}_{>0}(\mathfrak{g})\subset{\mathcal U}(\mathfrak{g})$ of strictly positive degree is always an associative ideal.

\vspace{3mm}
\noindent{\small\textbf{Example (Abelian Lie algebra):} Any vector space $V$ can be endowed with a structure of Lie algebra with the trivial Lie bracket. The universal enveloping algebra of the Abelian Lie algebra $V$ is isomorphic to the symmetric algebra of $V$:
${\mathcal U}(V)\,\cong\,\odot(V)$\,.
}

\vspace{3mm}
\noindent{\small\textbf{Remark:} When dealing with Lie-Rinehart algebras $\mathfrak{L}$ in the next subsection, it will be important to distinguish explicitly the universal enveloping algebra ${\mathcal U}_{\mathbb K}(\mathfrak{L})$ of the Lie algebra $\mathfrak{L}$ over the field $\mathbb K$ from the universal enveloping algebra ${\mathcal U}_{\mathcal A}(\mathfrak{L})$ of the Lie-Rinehart algebra $\mathfrak{L}$ over the commutative algebra $\mathcal A$. In the former case, the Lie-Rinehart algebra is simply seen as a Lie algebra (over the field $\mathbb K$) while, in the latter case, the richer structure of Lie-Rinehart algebra (over the commutative algebra $\mathcal A$) is taken into account (see Subsection \ref{UEALR}).
}

\subsubsection{Lie-Rinehart algebras and their associated bimodules}

To any Lie-Rinehart algebra $\mathfrak{L}$ over a commutative algebra $\mathcal A$, with anchor $\rho$, is
associated another Lie-Rinehart algebra over $\mathcal A$: the semidirect sum ${\mathfrak{B}}=\mathfrak{L}\inplus_\rho\mathfrak{A}$ of the Lie-Rinehart algebra $\mathfrak{L}$ and the commutator algebra $\mathfrak{A}$.

\vspace{3mm}
\noindent{\small\textbf{Proof:} Indeed, the anchor $\rho:\mathfrak{L} \to \mathfrak{der}({\mathcal A})$ of any Lie-Rinehart algebra $\mathfrak{L}$ over a commutative algebra $\mathcal A$ is a representation of the Lie algebra $\mathfrak{L}$ on the associative algebra $\mathcal A$. The corresponding Abelian Lie algebra $\mathfrak{A}$ can be endowed trivially with a structure of $\mathcal A$-Lie algebra (with trivial anchor and bracket). Therefore, the anchor of  $\mathfrak{L}$ provides a representation $\rho:\mathfrak{L}\to\mathfrak{der}(\mathfrak{A})$ of the Lie-Rinehart algebra $\mathfrak{L}$ on the $\mathcal A$-Lie algebra $\mathfrak{A}$. This allows to define the semidirect sum ${\mathfrak{B}}=\mathfrak{L}\inplus_\rho\mathfrak{A}$.
By construction, the adjoint representation of the subalgebra $\mathfrak{L}\subset\mathfrak{B}$ on 
the Abelian ideal ${\mathfrak{A}}$ is defined by the anchor $\rho$ of $\mathfrak{L}$, \textit{i.e.} $[\hat{X},\hat{f}]_{{}_{\mathfrak{B}}}:=\hat{X}[f]$ for any $\hat{X}\in \mathfrak{L}$ and $f\in\mathcal A$. 
The left $\mathcal A$-module structure of ${\mathfrak{B}}$ is defined in the obvious way $g\cdot(f\oplus\hat{X}):=(g\cdot f)\oplus(g\cdot\hat{X})$.
\qed
}
\vspace{3mm}

What is remarkable is that this Lie-Rinehart algebra ${\mathfrak{B}}={\mathfrak{A}}\niplus \mathfrak{L}$ has a richer structure than the original Lie-Rinehart algebra $\mathfrak{L}$ in the sense that it is not only a left $\mathcal A$-module but also an $\mathcal A$-bimodule, where the right $\mathcal A$-module structure is defined via the action
\be\label{canactLA}
(f\oplus \hat{X})\bullet g\,:=\, \big(\,f\cdot g+\hat{X}[g]\,\big)\oplus\big(\,g\cdot\hat{X}\,\big)\,.
\ee
From this perspective, the anchor of $\mathfrak{L}$ can be seen as the structure that relates the left and right $\mathcal A$-module structures of ${\mathfrak{A}}\niplus \mathfrak{L}$. If the left and right actions are written by the same symbol $\circ$ in an operatorial form, then the relation \eqref{canactLA} can be written in the more balanced form $(\hat{f}\oplus \hat{X})\circ \hat{g}\,:=\, \big(\,\hat{f}\circ \hat{g}+\hat{X}[g]\,\big)\oplus\big(\,\hat{g}\circ\hat{X}\,\big)$, which already suggests its later interpretation in the universal enveloping algebra as arising from a commutator of an associative product. 

\vspace{3mm}
\noindent{\small\textbf{Example (Lie-Rinehart algebra of smooth vector fields)\,:} In the case when $\mathcal A$ and $\mathfrak{L}$ are respectively the structure algebra ${\mathcal C}^\infty(M)$ and the Lie algebra ${\mathfrak{X}}(M)$ of vector fields on a manifold $M$, the associated Lie-Rinehart algebra ${\mathfrak B}={\mathfrak{A}}\niplus \mathfrak{L}$ is isomorphic to the Lie-Rinehart algebra ${\mathfrak{D}}^1(M)\cong {\mathcal C}^\infty(M)\niplus{\mathfrak{X}}(M)$ of first-order differential operators on $M$. 
}

\subsubsection{Universal enveloping algebras of Lie-Rinehart algebras}\label{UEALR}

The \textbf{universal enveloping algebra} ${\mathcal U}_{\mathcal A}(\mathfrak{L})$ \textbf{of a Lie-Rinehart algebra} $\mathfrak{L}$ \textbf{over the commutative algebra} $\mathcal A$ can be defined (see \textit{e.g.} \cite{Moerdijk} and refs therein) as the quotient
of the universal enveloping algebra ${\mathcal U}_{\mathbb K}({\mathfrak{B}})$ of the Lie algebra ${\mathfrak{B}}=\mathfrak{A}\niplus \mathfrak{L}$ by the two-sided ideal generated by the left action of $\mathcal A$ on ${\mathfrak{B}}$, \textit{i.e.} by the associative ideal spanned by elements proportional to $g\cdot(f\oplus\hat{X})-(g\cdot f)\oplus(g\cdot\hat{X})$ for some $f,g\in\mathcal A$ and $\hat{X}\in \mathfrak{L}$.

\vspace{3mm}
\noindent{\small\textbf{Example (Smooth differential operators)\,:} 
The almost-commutatative algebra ${\mathcal D}(M)$ of differential operators  on $M$ is the universal enveloping algebra of the Lie-Rinehart algebra ${\mathfrak{X}}(M)$ of vector fields on $M$, \textit{i.e.} 
\be
{\mathcal U}_{C^\infty(M)}\big({\mathfrak{X}}(M)\big)\cong{\mathcal D}(M)\,.
\ee
}

\vspace{1mm}
\noindent{\small\textbf{Example (Free modules generated by Lie algebras)\,:} Consider the simplest examples of $\mathcal A$-Lie algebras:  free ${\mathcal A}$-modules generated by a Lie algebra $\mathfrak{g}$. The universal enveloping algebra of such an $\mathcal A$-Lie algebra $\mathfrak{L}={\mathcal A}\otimes\mathfrak{g}$ is the free ${\mathcal A}$-modules generated by the universal enveloping algebra of the Lie algebra $\mathfrak{g}$, 
\be
{\mathcal U}_{\mathcal A}({\mathcal A}\otimes\mathfrak{g})\cong{\mathcal A}\otimes{\mathcal U}_{\mathbb K}(\mathfrak{g})\,.
\ee
}

\vspace{3mm}
\noindent{\small\textbf{Example (Invariant differential operators)\,:} 
Consider a \textbf{Klein geometry}, \textit{i.e.}, a pair made of a Lie group $G$ and a closed Lie subgroup $H\subset G$ such that the coset space $G/H$ is connected. It defines a principal $H$-bundle $G$ over $G/H$ whose Atiyah algebroid is the vector bundle $\frac{TG}{H}$ over $G/H$. Its global sections are $H$-invariant vector fields on $G$. They span the Atiyah algebra $\mathfrak{X}(G)^H=\Gamma(TG/H)$. 
The universal enveloping algebra of the Atiyah algebra of such a Klein geometry $H\subset G$ is spanned by $H$-invariant differential operators on $G$ \cite[Example 4.26]{Bekaert:2022dlx}
\be
	{\cal U}_{C^\infty(G/H)}\big(\mathfrak{X}(G)^H\big)\simeq{\cal D}(G)^H\simeq {\cal U}_{C^\infty(G)}\big(\mathfrak{X}(G)\big)^H.
\ee
}

\vspace{3mm}
\noindent{\small\textbf{Counter-example (Covariant differential operators on a module)\,:} Consider an $\mathcal A$-module $\textsc{V}$. The tensor algebra $\otimes_{\mathcal A}(\textsc{V})=\oplus_{r\in\mathbb N}\otimes^r_{\mathcal A}\textsc{V}$ is an $\mathcal A$-algebra, $\mathbb{N}$-graded by the rank of tensors.
A covariant derivative on an $\mathcal A$-module $\textsc{V}$ can be defined equivalently as a derivation of the tensor algebra $\otimes_{\mathcal A}\textsc{V}$ preserving the rank of tensors. Consider the algebra ${\mathcal D}_{\mathcal A}(\otimes_{\mathcal A}\textsc{V})=\oplus_{q\in\mathbb Z}{\mathcal D}_q(\otimes_{\mathcal A}\textsc{V})$ of differential operators on the tensor algebra $\otimes_{\mathcal A}(\textsc{V})$. It is $\mathbb{Z}$-graded: an element of ${\mathcal D}_q(\otimes_{\mathcal A}\textsc{V})$ increases the tensor rank by $q$.
A differential operator on the tensor algebra $\otimes_{\mathcal A}(\textsc{V})$ preserving the rank of tensors will be called a \textbf{covariant differential operator on the} $\mathcal A$-\textbf{module} $\textsc{V}$ since they are somehow the higher-order generalisation of covariant derivatives.
The almost-commutative subalgebra ${\mathcal D}_0(\otimes_{\mathcal A}\textsc{V})$ spanned by all covariant differential operators on the $\mathcal A$-module $\textsc{V}$ will be denoted ${\mathcal CD}_{\mathcal A}(\textsc{V})$.
In general, it is \textit{not} isomorphic to 
the universal enveloping algebra ${\mathcal U}_{\mathcal A}(\,\mathfrak{cder}_{\mathcal A}(\textsc{V})\,)$ of the Atiyah algebra $\mathfrak{cder}_{\mathcal A}(\textsc{V})$ of covariant derivatives \cite[Example 4.28]{Bekaert:2022dlx}. However, if the first-order covariant differential operators generate the whole algebra of covariant differential operators, then it is isomorphic to a quotient of the universal enveloping algebra.
}

\vspace{3mm}
Remember that the universal enveloping algebra ${\mathcal U}(\mathfrak{g})$ of a Lie algebra $\mathfrak{g}$ is never simple. This remains true for the universal enveloping algebra ${\mathcal U}_{\mathcal A}(\mathfrak{L})$ of an $\mathcal A$-Lie algebra $\mathfrak{L}$.
However, for a Lie-Rinehart algebra $\mathfrak{L}$ with non-trivial anchor the universal enveloping algebra ${\mathcal U}_{\mathcal A}(\mathfrak{L})$ may be simple. 

\vspace{3mm}
\noindent{\small\textbf{Example (Lie-Rinehart algebra of polynomial differential operators)\,:} The Weyl algebra ${\mathcal D}(A)$ of polynomial differential operators on the affine space $A$, modeled on the vector space $V$, is simple although it is isomorphic to the universal enveloping algebra of the Lie-Rinehart algebra $\mathfrak{der}(\odot V^*)$ of polynomial vector fields on $A$,
\be
{\mathcal U}_{\odot V^*}\big(\mathfrak{der}(\odot V^*)\,\big)\cong{\mathcal D}(A)\,.
\ee
}

\subsubsection{Poincar\'e-Birkhoff-Witt theorem}

In order to identify concretely the universal enveloping algebra of a Lie-Rinehart algebra, the most important result to know is the generalisation \cite{Rinehart} by Rinehart of the Poincar\'e-Birkhoff-Witt theorem: if $\mathfrak{L}$ is a projective left $\mathcal A$-module, then the universal enveloping algebra ${\mathcal U}_{\mathcal A}(\mathfrak{L})$ is an almost-commutative algebra whose graded algebra is isomorphic to the symmetric algebra of $\mathfrak{L}$ over $\mathcal A$, 
\be
\text{gr}\,{\mathcal U}_{\mathcal A}(\mathfrak{L})\,\cong\,\odot_{\mathcal A}(\mathfrak{L})\,.
\ee
It is important to emphasise that the symmetric algebra $\odot_{\mathcal A}(\mathfrak{L})$ over ${\mathcal A}$ is much smaller than the usual symmetric algebra $\odot_{\mathbb K}(\mathfrak{L})$ over ${\mathbb K}$, because the former takes into account the full ${\mathcal A}$-linearity of the corresponding tensor product (while the latter only takes into account its ${\mathbb K}$-linearity).

\vspace{3mm}
\noindent{\small\textbf{Example (Smooth differential operators)\,:} 
The almost-commutatative algebra ${\mathcal D}(M)$ of differential operators on $M$ is the universal enveloping algebra of the Lie-Rinehart algebra ${\mathfrak{X}}(M)$ of vector fields on $M$. From the generalised Poincar\'e-Birkhoff-Witt theorem, one recovers that the classical limit of the almost-commutative algebra ${\mathcal D}(M)$ of differential operators  on $M$ is isomorphic to the Schouten algebra ${\mathcal S}(M)\cong\Gamma(\odot TM)$ of principal symbols on $M$, \textit{i.e.} 
\be
\text{gr}\,{\mathcal D}(M)\cong\odot_{{}_{C^\infty(M)}}\big({\mathfrak{X}}(M)\big)\cong\mathcal{S}(M)\,.
\ee
}

\vspace{3mm}
\noindent{\small\textbf{Example (Polynomial vs formal differential operators)\,:} The Grothendieck algebra of differential operators acting on the commutative algebra $\mathcal{A}=\odot(V^*)$ of polynomials on the affine space $A$ (respectively, the commutative algebra $\mathcal{A}=\overline{\odot}(V^*):=\odot(V)^*$ of formal power series at the origin of the vector space $V$)  is the universal enveloping algebra of the Lie-Rinehart algebra $\mathfrak{der}({\mathcal A})$ of polynomial (respectively, formal) vector fields $\hat{X}=X^{a}(y)\,\partial_a$: 
\be
{\mathcal U}_{\mathcal A}\big(\mathfrak{der}({\mathcal A})\,\big)={\mathcal D}({\mathcal A})\qquad\text{for}\quad{\mathcal A}\,\,=\,\,\text{either}\,\,\odot(V^*)\,\,\text{or}\,\,\overline{\odot}(V^*)\,.
\ee
The Poincar\'e-Birkhoff-Witt theorem leads to the isomorphism
\be\label{PBWA}
\text{gr}\,{\mathcal D}({\mathcal A})\,\cong\,{\mathcal A}\,\otimes\,\odot(V)\qquad\text{for}\quad{\mathcal A}\,\,=\,\,\text{either}\,\,\odot(V^*)\,\,\text{or}\,\,\overline{\odot}(V^*)\,,
\ee
since $\mathfrak{der}({\mathcal A})\cong {\mathcal A}\otimes V$ as vector spaces.
The isomorphism \eqref{PBWA} is the property that the commutative algebra $\text{gr}\,{\mathcal D}({\mathcal A})$ of symbols
\be
X=\sum\limits_{r=0}^k X^{a_1\cdots a_r}(y)\,p_{a_1}\cdots p_{a_r}
\ee
of differential operators 
\be
\hat{X}=\sum\limits_{r=0}^k X^{a_1\cdots a_r}(y)\,\partial_{a_1}\cdots\partial_{a_r}
\ee
is a free $\mathcal A$-module with all monomials $p_{a_1}\cdots p_{a_r}$ as holonomic basis, \textit{i.e.} the corresponding components $X^{a_1\cdots a_r}(y)$ are polynomials (vs formal power series).
In particular, the symbols
of polynomial differential operators are polynomials functions on the cotangent bundle $T^*V=V\oplus V^*$, in agreement with the isomorphism
\be\label{clWeyl}
\odot(V\oplus V^*)\,\cong\,\odot(V)\,\otimes\,\odot(V^*)
\ee
}

\subsubsection{Universality property}

An equivalent (but more abstract) definition of the universal enveloping algebra ${\mathcal U}_{\mathcal A}(\mathfrak{L})$ of the Lie-Rinehart algebra $\mathfrak{L}$ over the commutative algebra $\mathcal A$ is by the following universality property (see \textit{e.g.} \cite{Moerdijk,Huebschmann}).\footnote{Apart from the well-known constructions of Rinehart \cite{Rinehart} and Huebschmann \cite{Huebschmann}, the universal enveloping algebra of a Lie-Rinehart algebra admits other (equivalent) realisations (see \textit{e.g.} \cite{Kaoutit,Saracco} and refs therein).}

Let $\mathfrak{L}$ be a Lie-Rinehart algebra over $\mathcal A$ with $\cdot$ denoting the left action of $\mathcal A$ on $\mathfrak{L}$. Let $\cal U$ be an associative algebra with product denoted by $\circ$ and with commutator algebra denoted by $\mathfrak{U}$. 
If 
\be
a\,:\,{\mathcal A}\to{\mathcal U}\,:\,f\mapsto a(f)
\ee
is a morphism of associative algebras and 
\be
\ell\,:\,\mathfrak{L}\to\mathfrak{U}\,:\,\hat{X}\mapsto\ell({X})
\ee
is a morphism of Lie algebras, such that they satisfy the compatibility conditions ($\forall f\in\mathcal A$, $\forall {X}\in\mathfrak{L}$)\,:
\be\label{conduniv1}
\ell(f\cdot{X})\,=\,a(f)\circ\ell({X})\,,
\ee
and
\be\label{conduniv2}
\ell({X})\,\circ\,a(f)\,-\,a(f)\,\circ\,\ell(X)\,=\,a\big(\,\hat{X}[f]\,\big)\,,
\ee
then there exists a unique extension 
\be
u\,:\,{\mathcal U}_{\mathcal A}(\mathfrak{L})\to{\mathcal U}\,,\qquad u|_{\mathcal A}=a\,,\quad u|_{\mathfrak{L}}=\ell\,,
\ee
which is a morphism of associative algebras. 

An important corollary of this universality property is that there is a one-to-one correspondence between modules of a Lie-Rinehart algebras $\mathfrak{L}$ over $\mathcal A$ and modules of its universal enveloping algebra ${\mathcal U}_{\mathcal A}(\mathfrak{L})$.
In fact, if a vector space $\textsc V$ carries a representation of a Lie-Rinehart algebra $\mathfrak{L}$ over $\mathcal A$ then it is both a left $\mathcal A$-module $\textsc V$ (\textit{i.e.} there is a morphism $a:{\mathcal A}\to\text{End}_{\mathcal A}(\textsc{V})$ of associative algebras) and a left
$\mathfrak{L}$-module (\textit{i.e.} there is a morphism $\nabla:\mathfrak{L}\to\mathfrak{cder}_{\mathcal A}(\textsc{V})$ of Lie-Rinehart algebras).
Setting $\ell=\nabla$ and ${\mathcal U}={\mathcal D}_{\mathcal A}(\textsc{V})$, one finds that there exists a unique extension 
${\mathcal U}(\nabla):{\mathcal U}_{\mathcal A}(\mathfrak{L})\to{\mathcal D}_{\mathcal A}(\textsc{V})$ as a morphism of associative algebras. This makes $\textsc V$ a left ${\mathcal U}_{\mathcal A}(\mathfrak{L})$-module.

Note that the extension of a faithful representation $\nabla:\mathfrak{L}\hookrightarrow \mathfrak{cder}_{\mathcal A}(\textsc{V})$ of a Lie-Rinehart algebra to a 
representation ${\mathcal U}(\nabla):{\mathcal U}_{\mathcal A}(\mathfrak{L})\to {\mathcal D}_{\mathcal A}(\textsc{V})$ of the universal enveloping algebra may not be faithful. However, the representation of the almost-commutative algebra ${\mathcal U}_{\mathcal A}(\mathfrak{L})\,/\,\text{Ker}\,{\mathcal U}(\nabla)$ defined as the quotient of the universal enveloping algebra ${\mathcal U}_{\mathcal A}(\mathfrak{L})$ by the kernel of ${\mathcal U}(\nabla)$ will be faithful by construction. This quotient will be called the \textbf{enveloping algebra of the Lie-Rinehart algebra} $\mathfrak{L}$ \textbf{associated to the faithful representation} $\nabla$. In this sense, any faithful representation of a Lie-Rinehart algebra (such as a flat connection) can be lifted to a faithful representation of the corresponding enveloping algebra. And, conversely, any (faithful) representation of an almost-commutative algebra restricts to a (faithful) representation of a Lie-Rinehart algebra.

\subsubsection{Almost-commutative algebras and associative algebroids}

Consider an almost-commutative algebra $\cal U$ and let us denote by $\mathfrak{U}$ its commutator algebra. 
Recall that the component of degree zero is a commutative algebra ${\mathcal U}_0$ and that the component of degree one is a Lie-Rinehart algebra $\mathfrak{U}_1$ over ${\mathcal U}_0$.
Furthermore,
the quotient $\mathfrak{U}_1/\mathfrak{U}_0\subset\text{gr}\,\mathfrak{U}$ is a Lie-Rinehart subalgebra of the grade one component of the classical limit $\text{gr}\,{\mathcal U}$. The Lie-Rinehart algebra $\mathfrak{U}_1$ over ${\mathcal U}_0$ is the extension \eqref{U0U1quotient} of the Lie-Rinehart subalgebra $\mathfrak{U}_1/\mathfrak{U}_0\subset\text{gr}\,\mathfrak{U}$  by the Abelian Lie-Rinehart subalgebra $\mathfrak{U}_0\subset\mathfrak{U}$.

Another corollary of the universality property is that, for any almost-commutative algebra $\cal U$, there exists a morphism of associative algebras from the universal enveloping algebra $\mathcal{U}_{{}_{{\mathcal U}_0}}(\mathfrak{U}_1/\mathfrak{U}_0)$ of the Lie-Rinehart algebra $\mathfrak{U}_1/\mathfrak{U}_0$ over ${\mathcal U}_0$ to the almost-commutative algebra $\cal U$ (see \textit{e.g.} \cite[Section 2.1]{Martinez} for more details).
In this sense, ``almost-commutative'' algebras could be called ``associative-Rinehart'' algebras, since almost-commutative algebras are to Lie-Rinehart algebras what associative algebras are to Lie algebras.\footnote{This statement can even be made precise in functorial language. The ``commutator'' functor associating a Lie-Rinehart algebra to any almost-commutative algebra is right-adjoint to the ``universal enveloping'' functor associating an almost-commutative algebra to any Lie-Rinehart algebra \cite[Proposition 2.9]{Martinez}.}
Accordingly, an almost-commutative algebra $\mathcal{U}$ whose degree zero component $\mathcal{U}_0$ is the structure algebra $C^\infty(M)$ of a manifold $M$ and such that each component $\mathcal{U}_k$ is locally-free of finite-rank, could be called an \textbf{associative algebroid} over $M$. In particular, an associative algebroid is the space of sections of a filtered vector bundle over $M$ with two important vector sub-bundles: the unit bundle $M\times\mathbb R$ at degree zero, and a Lie algebroid $\mathbb{A}$ at degree one. As argued in the introduction, it is tempting to speculate that associative algebroids should be the proper arena for discussing geometrically higher-spin gauge symmetries and connections.

An almost-commutative algebra $\cal U$ generated by its component ${\mathcal U}_1$ of degree one will be called an \textbf{enveloping algebra of the Lie-Rinehart algebra} $\mathfrak{U}_1/\mathfrak{U}_0$ over the commutative algebra ${\mathcal U}_0$. Another corollary of the universality property is that any enveloping algebra $\cal U$ of a Lie-Rinehart algebra $\mathfrak{L}$ over the commutative algebra ${\mathcal A}$ is isomorphic to a quotient of the universal enveloping algebra ${\mathcal U}_{\mathcal A}(\mathfrak{L})$ of the Lie-Rinehart algebra $\mathfrak{L}$ over $\mathcal A$. In fact, the morphism ${\mathcal U}_{\mathcal A}(\mathfrak{L})\to\cal U$ of associative algebras is surjective since $\cal U$ is generated by its component of order one ${\mathcal U}_1$, therefore one has a short exact sequence of associative algebra morphisms
\be
0\to {\mathcal I}\stackrel{i}{\hookrightarrow}{\mathcal U}_{\mathcal A}(\mathfrak{L})\stackrel{\pi}{\twoheadrightarrow}{\mathcal U}\to 0
\ee
where the associative ideal ${\mathcal I}$ of ${\mathcal U}_{\mathcal A}(\mathfrak{L})$ is the kernel of $\pi$.

\subsection{Weyl algebra as enveloping algebra of Heisenberg algebra}

Let $V$ be a finite-dimensional vector space.
Its cotangent bundle $T^*V\cong V\oplus V^*$ is endowed with a canonical symplectic two-form $\Omega$ defined by $\Omega(v\oplus\alpha,w\oplus\beta)=\alpha(w)-\beta(v)$ for all $v,w\in V$ and $\alpha,\beta\in V^*$. Conversely, any finite-dimensional symplectic vector space $W$ admits a choice of polarisation $W=V\oplus V^*$.

\subsubsection{Heisenberg group and algebra}

Obviously, the vector space $V\oplus V^*$ can be seen as an additive Abelian Lie group.  The \textbf{Heisenberg group} $H(V)$ is a nontrivial central extension
\be
0\to {\mathbb K}\,\hookrightarrow\,H(V)\,\twoheadrightarrow\, V\oplus V^* \to0\,.\label{shortexactHeinsenberg}
\ee
of the Abelian Lie group $V\oplus V^*$ by the Abelian Lie group ${\mathbb K}$.
It is the vector space $V\oplus V^*\oplus{\mathbb K}$ endowed with the product 
\ba
&&(v,\alpha,t)\cdot (w,\beta,u)= \Big(\,v+w\,,\alpha+\beta\,,\,t+u+\alpha(w)-\beta(v)\,\Big)\,,\\
&&\qquad \forall\, v,w\in V\,,\quad\forall\,\alpha,\beta\in V^*,\quad\forall\, t,u\in{\mathbb K}\,.\nonumber
\ea
The Heisenberg group $H(V)$ is a non-Abelian Lie group whose center is ${\mathbb K}$.

The \textbf{Heisenberg algebra} $\mathfrak{h}(V)$ is the Lie algebra of the Heisenberg group $H(V)$. 
It is the vector space $V\oplus V^*\oplus{\mathbb K}$ endowed with the Lie bracket
\ba
&&\big[\,(v,\alpha,t)\,,\,(w,\beta,u)\,\big]\,=\, \big(\,0,0,\,\alpha(w)-\beta(v)\,\big)\,,\\
&&\qquad \forall\, v,w\in V\,,\quad\forall\,\alpha,\beta\in V^*,\quad\forall\, t,u\in{\mathbb K}\,.\nonumber
\ea
Given a basis $\{e_a\}$ of the vector space $V$, the latter becomes isomorphic to ${\mathbb K}^n$ in which case the Heisenberg group (respectively, algebra) is often denoted by physicists $H_{2n}$ (respectively, $\mathfrak{h}_{2n}$). 
Let $\texttt{c}$ denote the central element of $\mathfrak{h}(V)$ corresponding to the unit element $1\in\mathbb K$. In the basis $\{e_a,e^{*b},\texttt{c}\}$ of $\mathfrak{h}_{2n}$, the only nontrivial Lie brackets are given by $[e^{*b},e_a]=\delta_a^b\texttt{c}$.

\subsubsection{Unitary irreducible representations}

The theorem of Stone and von Neumann asserts (respectively, implies) that all unitary irreducible representations of the real Heisenberg group $H(V)$ (respectively, of the real Heisenberg algebra $\mathfrak{h}(V)$\,) which are not trivial on the center ${\mathbb R}$ are unitarily equivalent (up to a scale, i.e. up to a rescaling of the eigenvalue of the central element).

By Schur's lemma, all unitary irreducible modules of the Heisenberg algebra $\mathfrak{h}(V)$ are eigenspaces of the central element $\texttt{c}$. All unitary irreducible modules of the Heisenberg algebra $\mathfrak{h}(V)$ for non-vanishing real eigenvalue look exactly the same, so one may take $1$ as eigenvalue.
This faithful representation of the Lie algebra $\mathfrak{h}(V)$ extends to a representation of its universal enveloping algebra ${\mathcal U}\big(\,\mathfrak{h}(V)\,\big)$ which is \textit{not} faithful. The associative ideal $(\texttt{c}-1)\,{\mathcal U}\big(\,\mathfrak{h}(V)\,\big)$ is the annihilator of the corresponding unitary irreducible $\mathfrak{h}(V)$-module. The Weyl algebra ${\mathcal D}(A)$ is isomorphic to the quotient 
\be
{\mathcal U}\big(\mathfrak{h}(V)\big)\,/\,(\texttt{c}-1){\mathcal U}\big(\mathfrak{h}(V)\big)\cong {\mathcal D}(A)
\ee 
of the universal enveloping algebra ${\mathcal U}\big(\mathfrak{h}(V)\big)$ of the Heisenberg algebra $\mathfrak{h}(V)$ by the primitive ideal $(\texttt{c}-1){\mathcal U}\big(\mathfrak{h}(V)\big)$. In other words, the Weyl algebra is isomorphic to the enveloping algebra of the Heisenberg algebra associated to one of its representation on a unitary irreducible module, non-trivial on the centre.

The classical limit $\text{gr}\,{\mathcal D}(A)$ of the Weyl algebra (seen as an almost-commutative algebra) is isomorphic to the Schouten algebra 
\eqref{clWeyl}
of polynomial functions on the cotangent space $T^*V\cong V\oplus V^*$.
This algebra \eqref{clWeyl} of polynomial symbols is isomorphic to the quotient of the universal enveloping algebra ${\mathcal U}\big(\mathfrak{h}(V)\big)$ of the Heisenberg algebra by the associative ideal $\texttt{c}\,{\mathcal U}\big(\mathfrak{h}(V)\big)$,
\be
\odot(V\oplus V^*)\cong{\mathcal U}\big(\mathfrak{h}(V)\big)\,/\,\texttt{c}\,{\mathcal U}\big(\mathfrak{h}(V)\big)\,.
\ee
In fact, this quotient amounts to take the classical limit where position and momenta commute with each other.

\vspace{5mm}\begin{figure}
	\begin{framed}
		\begin{center}
			\textbf{The many faces of Weyl algebras}
		\end{center}
		
		\noindent
		Consider an affine space $A$ modeled on a vector space $V$.
		The Weyl algebra ${\mathcal D}(A)$ can be defined in various equivalent ways as:
		
		\noindent$\bullet$ the Grothendieck algebra ${\mathcal D}(\odot V^*)$ of the commutative algebra of polynomial functions on $A$,
		
		\noindent$\bullet$ the universal enveloping algebra ${\mathcal U}_{\odot V^*}\big(\mathfrak{der}(\odot V^*)\,\big)$ of the Lie-Rinehart algebra of polynomial vector fields on $A$,
		
		\noindent$\bullet$ the enveloping algebra $\frac{{\mathcal U}\big(\mathfrak{h}(V)\big)}{(\texttt{c}-1){\mathcal U}\big(\mathfrak{h}(V)\big)}$ of the Heisenberg algebra associated to one of its representation on a unitary irreducible module non-trivial on the centre,
	\end{framed}
\end{figure}

\subsection{Universal enveloping algebras of semidirect sums}

For the sake of simplicity of the discussion, let us focus first on the example of Lie algebras over a field $\mathbb{K}$. 

The universal enveloping algebra ${\mathcal U}(\mathfrak{g})$ of a semidirect sum 
\be
\mathfrak{g}=\mathfrak{i}\niplus\mathfrak{h}
\ee
of the Lie ideal $\mathfrak{i}\subset\mathfrak{g}$ and the Lie subalgebra $\mathfrak{h}\subset\mathfrak{g}$ is isomorphic to the smash product of the respective universal enveloping algebras ${\mathcal U}(\mathfrak{i})$ and ${\mathcal U}(\mathfrak{h})$  \cite[Subsection 1.7.11]{McConnell},
\be\label{isomsemicrossed}
{\mathcal U}(\mathfrak{i}\niplus\mathfrak{h})\,\cong\,{\mathcal U}(\mathfrak{i})\rtimes{\mathcal U}(\mathfrak{h})\,,
\ee
where the action of $\mathfrak{h}$ on ${\mathcal U}(\mathfrak{i})$ arises via the Leibnitz rule from the representation of $\mathfrak{h}$ on $\mathfrak{i}$.
This result admits a generalisation \cite{BCM,Montgomery} to the case of a linearly-split extension\footnote{In other words, the arrows in the splitting $0\leftarrow \mathfrak{i}\twoheadleftarrow\mathfrak{g}\hookleftarrow\mathfrak{h}\leftarrow 0$ are morphisms of vector spaces only.} $\mathfrak{g}$ of the Lie algebra $\mathfrak{h}$ by the ideal $\mathfrak{i}$, in which case the symbol $\niplus$ stands for the ``curved'' semidirect sum \cite[Definition 1.7]{Bekaert:2022dlx} while the symbol  $\rtimes$ in \eqref{isomsemicrossed} stands for the ``cross'' product \cite[Definition 4.1]{BCM}.
The abstract definitions of the smash and cross products $\rtimes$ will not be reviewed here (because it involves some concepts in bialgebra theory that are beyond the scope of the present text).\footnote{For those interested, see \textit{e.g.} \cite{Hazewinkel:2010} for a thorough introduction to bialgebras, Hopf algebras, etc.} Anyway, in order to understand the meaning of \eqref{isomsemicrossed}, it is enough to appreciate that the generalised Poincar\'e-Birkhoff-Witt theorem implies that
\be
\text{gr}\,{\mathcal U}(\mathfrak{i}\niplus\mathfrak{h})\,\,\cong\,\,\odot(\mathfrak{i}\oplus\mathfrak{h})\,\,\cong\,\,\odot(\mathfrak{i})\,\otimes\,\odot(\mathfrak{h})\,,
\ee
where the associated graded algebra is with respect to both filtrations, \textit{i.e.} of ${\mathcal U}(\mathfrak{i})$ and of ${\mathcal U}(\mathfrak{h})$. 
More concretely, there is a natural choice of ordering for ${\mathcal U}(\mathfrak{i}\niplus\mathfrak{h})$: the ``normal'' ordering where the dependence in $\mathfrak{i}$ is factored on the left while the dependence on $\mathfrak{h}$ is factored on the right. 
The product of two normal-ordered elements of ${\mathcal U}(\mathfrak{i}\niplus\mathfrak{h})$ is not any more normal-ordered in general.
The normal ordering requires to recursively compute commutators of the form $[{\mathcal U}(\mathfrak{h}),{\mathcal U}(\mathfrak{i})]$. 
In some sense, the abstract notion of smash product is simply a way to formalise the systematic calculus (use of Leibnitz rule, etc) involved with the normal ordering of such expressions.

\vspace{3mm}
\noindent{\small\textbf{Example (Direct sum)\,:} The universal enveloping algebra ${\mathcal U}(\mathfrak{g})$ of a direct sum 
\be
\mathfrak{g}=\mathfrak{h_1}\oplus\mathfrak{h_2}
\ee
of two Lie algebra $\mathfrak{h}_1$ and $\mathfrak{h}_2$ is isomorphic to the tensor product of the respective universal enveloping algebras ${\mathcal U}(\mathfrak{h}_1)$ and ${\mathcal U}(\mathfrak{h}_2)$,
\be
{\mathcal U}(\mathfrak{h}_1\oplus\mathfrak{h}_2)\,\cong\,{\mathcal U}(\mathfrak{h}_1)\otimes{\mathcal U}(\mathfrak{h}_2)\,.
\ee
This obvious result corresponds to the isomorphism \eqref{isomsemicrossed} for the case of a trivial representation.
}
\vspace{3mm}

Interestingly, the factorisation \eqref{isomsemicrossed} admits a generalisation for Lie-Rinehart algebras \cite{Bekaert:2022dlx}: for any given curved (respectively, flat) connection, that is, a linear (respectively, Lie-Rinehart) splitting of a Lie-Rinehart algebra extension (\textit{i.e.} a generalised connection), a crossed (resp.~smash) product decomposition of the associated universal enveloping algebra is provided, and vice versa. 
As a geometric example for Lie algebroids, the associative algebra generated by the invariant vector fields on the total space of a principal bundle is described as a crossed product of the algebra generated by the vertical ones and the algebra of differential operators on the base. Such a factorisation can be thought as an alternative characterisation of an infinitesimal connection on a principal bundle.  Its interest for higher-spin geometry is that such an algebraic characterisation might admit natural generalisations adapted to the characterisation of higher-spin connections, \textit{e.g.} by relaxing in \cite[Theorem 3.10]{Bekaert:2022dlx} the condition that the coproduct (hence the filtration) of the universal enveloping algebra is preserved.

\pagebreak

\section*{Acknowledgments}

I would like to thank Damien Calaque for pointing to me (a long time ago) the relevance of Lie-Rinehart algebras for defining properly the universal enveloping algebra of the vector field Lie algebra. I am also very grateful to Niels Kowalzig and Paolo Saracco for our collaboration on the universal enveloping algebra of Lie-Rinehart algebra, from which I learned so much. Finally, I acknowledge Thomas Basile for his patient reading and useful comments on some early version of these notes.

The author acknowledge support of the Institut Henri Poincar\'e (UAR 839
CNRS-Sorbonne Université) and LabEx CARMIN (ANR-10-LABX-59-01) during his participation to  the trimester ``Higher Structures in Geometry
and Mathematical Physics'' held at the Institut Henri Poincar\'e (April-July 2023).



\end{document}